%% file: comix.tex
\documentclass[a4paper,fleqn,10pt]{article}
\input{text/global}

\begin{document}
%= fmffile ========================================
\begin{fmffile}{comix_fg}
\input{text/graphs}
%= title ==========================================
\mypreprint{IPPP/08/31\\DCPT/08/62\\[1mm]
  SLAC-PUB-13232\\[1.5mm]MCNET/08/08}
{\mytitle{Comix\symbolfootnote[1]{See
  \href{http://comix.freacafe.de}{http://comix.freacafe.de}
  for downloads and a detailed manual.}, 
  a new matrix element generator}
\myauthor{Tanju Gleisberg$^1$, Stefan H{\"o}che$^2$}
\myinstitute{$^1$ Stanford Linear Accelerator Center, 
  Stanford University, Stanford, CA 94309, USA\\
  $^2$ Institute for Particle Physics Phenomenology,
  Durham University, Durham DH1 3LE, UK}}
%= abstract =======================================
\myabstract{\input{text/abstract}}
%= introduction ===================================
\mysection{Introduction}
\pagenumbering{arabic}
\setcounter{page}{1}
\label{sec:intro}
\input{text/intro}
%= standard model =================================
\mysection{Recursive relations for tree-level 
  amplitudes in the Standard Model}
\label{sec:sm}
\input{text/sm}
%= me generation ==================================
\mysection{Matrix element generation in Comix}
\label{sec:me}
\input{text/me}

%= phase space ====================================
\mysection{Integration techniques in Comix}
\label{sec:ps}
\input{text/ps}
\clearpage
%= results ========================================
\mysection{Results}
\label{sec:results}
\input{text/results}

%= conclusions=====================================
\mysection{Conclusions}
\label{sec:conclusions}
\input{text/conclusions}
%= acknowledgements ===============================
\myssection{Acknowledgements}
\input{text/acknowledgements}
\clearpage
%= bibliography ===================================
\bibliographystyle{amsunsrt_mod}  
\bibliography{bibliography}
%= fmffile ========================================
\end{fmffile}
%= end ============================================
\end{document}

%% file: text/global.tex
\usepackage{amsmath}
\usepackage{amssymb}
\usepackage{hyperref}
\usepackage{array}
\usepackage{calc}
\usepackage{longtable}
\usepackage{axodraw}
\usepackage{pstricks}
\usepackage{graphicx}
\usepackage{xspace}
\usepackage{feynmp}
\usepackage{cite}
\usepackage{mcite}
\newcommand{\Comix}{C\protect\scalebox{0.8}{OMIX}\xspace}

\newcommand{\Amegic}{A\protect\scalebox{0.8}{MEGIC++}\xspace}

\newcommand{\Alpgen}{A\protect\scalebox{0.8}{LPGEN}\xspace}

\newcommand{\Haag}{H\protect\scalebox{0.8}{AAG}\xspace} 
\newcommand{\Vegas}{V\protect\scalebox{0.8}{EGAS}\xspace} 
\newcommand{\mypreprint}[1]{\begin{flushright}
  {\large #1}\end{flushright}\vspace*{2.5ex}}
\newcommand{\mytitle}[1]{{\flushleft\huge\sffamily\bfseries #1}}
\newcommand{\myauthor}[1]{{\flushleft\normalsize #1}}
\newcommand{\myinstitute}[1]{\flushleft{\small #1}\\\vspace*{5ex}}
\newcommand{\myabstract}[1]{\centerline{
  {\large\sffamily\bfseries Abstract:\hspace*{1ex}}\parbox[t]{
    \textwidth*5/6-\widthof{{\large\sffamily\bfseries Abstract:}
    \hspace*{1ex}}}{#1}}\vspace*{5ex}}

\newcommand{\mysection}[1]{\section{\Large\sffamily\bfseries #1}}
\newcommand{\mysubsection}[1]{\subsection{\large\sffamily\bfseries #1}}
\newcommand{\mysubsubsection}[1]{\subsubsection{\large\sffamily\bfseries #1}}
\newcommand{\myssection}[1]{\section*{\Large\sffamily\bfseries #1}}

\long\def\symbolfootnote[#1]#2{\begingroup%
\def\thefootnote{\fnsymbol{footnote}}\footnote[#1]{#2}\endgroup}

\newcommand{\rbr}[1]{\left( #1\right)}
\newcommand{\abr}[1]{\langle #1\rangle}
\newcommand{\cbr}[1]{\left\{ #1\right\}}
\newcommand{\sbr}[1]{\left[ #1\right]}
\newcommand{\done}{{\rm d}}

\newcommand{\mc}[1]{\mathcal{#1}}

\newcommand{\smallfeynmf}{
  \fmfset{thick}{1.25thin}
  \fmfset{arrow_len}{2mm}\fmfset{curly_len}{1.5mm}
  \fmfset{wiggly_len}{1.5mm}\fmfset{decor_size}{2mm}
  \fmfset{dot_len}{1mm}\fmfset{dot_size}{2thick}
  \fmfset{dash_len}{1.5mm}
  \unitlength=.5mm}

\newcommand{\myfigcaption}[2]{
  \refstepcounter{figure}
  \small{\bf Fig. \thefigure\hspace*{2ex}}
  \parbox[t]{#1-\widthof{\bf Fig. \thefigure\hspace*{4ex}}}{#2}
}
\newcommand{\myfigure}[3]{
  \begin{figure}[#1]
    \begin{center}
      #2\\\myfigcaption{\widthof{#2}}{#3}
    \end{center}
  \end{figure}
}
\newcommand{\mytable}[3]{
  \begin{table}[#1]
    \begin{center}
      #2\\\mytabcaption{\widthof{#2}}{#3}
    \end{center}
  \end{table}
}

%% file: text/graphs.tex
\newcommand{\dst}{\displaystyle}
\newcommand{\sst}{\scriptstyle}
% XXY vertex
\newcommand{\vxxs}[4]{\parbox{2cm}{\begin{center}
  \smallfeynmf\begin{fmfgraph*}(25,20)
    \fmftop{i,j}
    \fmfbottom{ij}
    \fmf{#1}{i,v1,j}
    \fmf{#2,tension=1.5}{v1,v2}
    \fmf{phantom,tension=5}{ij,v2}
    \fmfv{d.size=0,l.d=2.0,l.a=90,l=$\dst #3$}{i}
    \fmfv{d.size=0,l.d=2.0,l.a=90,l=$\dst #4$}{j}
    \fmfv{d.shape=circle,d.size=3,l.a=0}{v2}
  \end{fmfgraph*}\end{center}}}
\newcommand{\vxxl}[4]{\parbox{2cm}{\begin{center}
  \smallfeynmf\begin{fmfgraph*}(25,20)
    \fmftop{i,j}
    \fmfbottom{ij}
    \fmf{#1}{v1,i}
    \fmf{#2}{j,v1}
    \fmf{#1,tension=1.5}{v2,v1}
    \fmf{phantom,tension=5}{ij,v2}
    \fmfv{d.size=0,l.d=2.0,l.a=90,l=$\dst #3$}{i}
    \fmfv{d.size=0,l.d=2.0,l.a=90,l=$\dst #4$}{j}
    \fmfv{d.shape=circle,d.size=3,l.a=0}{v2}
  \end{fmfgraph*}\end{center}}}
\newcommand{\vxxr}[4]{\parbox{2cm}{\begin{center}
  \smallfeynmf\begin{fmfgraph*}(25,20)
    \fmftop{i,j}
    \fmfbottom{ij}
    \fmf{#2}{v1,i}
    \fmf{#1}{j,v1}
    \fmf{#1,tension=1.5}{v1,v2}
    \fmf{phantom,tension=5}{ij,v2}
    \fmfv{d.size=0,l.d=2.0,l.a=90,l=$\dst #3$}{i}
    \fmfv{d.size=0,l.d=2.0,l.a=90,l=$\dst #4$}{j}
    \fmfv{d.shape=circle,d.size=3,l.a=0}{v2}
  \end{fmfgraph*}\end{center}}}
\newcommand{\vxxf}[4]{\parbox{2cm}{\begin{center}
  \smallfeynmf\begin{fmfgraph*}(25,20)
    \fmftop{i,j}
    \fmfbottom{ij}
    \fmf{#1}{v1,i}
    \fmf{#1}{v1,j}
    \fmf{#2,tension=1.5}{v1,v2}
    \fmf{phantom,tension=5}{ij,v2}
    \fmfv{d.size=0,l.d=2.0,l.a=90,l=$\dst #3$}{i}
    \fmfv{d.size=0,l.d=2.0,l.a=90,l=$\dst #4$}{j}
    \fmfv{d.shape=circle,d.size=3,l.a=0}{v2}
  \end{fmfgraph*}\end{center}}}
% XXY interaction
\newcommand{\ixxs}[5]{\parbox{4cm}{\begin{center}
  \smallfeynmf\begin{fmfgraph*}(25,25)
    \fmftop{i,j}
    \fmfbottom{ij}
    \fmf{#1}{i,v1,j}
    \fmf{#2,tension=1.5}{v1,v2}
    \fmf{phantom,tension=2.5}{ij,v2}
    \fmfv{d.size=0,l.d=2.0,l.a=120,l=$\dst #3$}{i}
    \fmfv{d.size=0,l.d=2.0,l.a=60,l=$\dst #4$}{j}
    \fmfv{d.size=0,l.d=2.0,l.a=-90,l=$\dst #5$}{v2}
  \end{fmfgraph*}\end{center}}}
\newcommand{\ixxl}[5]{\parbox{4cm}{\begin{center}
  \smallfeynmf\begin{fmfgraph*}(25,25)
    \fmftop{i,j}
    \fmfbottom{ij}
    \fmf{#1}{v1,i}
    \fmf{#2}{j,v1}
    \fmf{#1,tension=1.5}{v2,v1}
    \fmf{phantom,tension=2.5}{ij,v2}
    \fmfv{d.size=0,l.d=2.0,l.a=120,l=$\dst #3$}{i}
    \fmfv{d.size=0,l.d=2.0,l.a=60,l=$\dst #4$}{j}
    \fmfv{d.size=0,l.d=2.0,l.a=-90,l=$\dst #5$}{v2}
  \end{fmfgraph*}\end{center}}}
\newcommand{\ixxr}[5]{\parbox{4cm}{\begin{center}
  \smallfeynmf\begin{fmfgraph*}(25,25)
    \fmftop{i,j}
    \fmfbottom{ij}
    \fmf{#2}{v1,i}
    \fmf{#1}{j,v1}
    \fmf{#1,tension=1.5}{v1,v2}
    \fmf{phantom,tension=2.5}{ij,v2}
    \fmfv{d.size=0,l.d=2.0,l.a=120,l=$\dst #3$}{i}
    \fmfv{d.size=0,l.d=2.0,l.a=60,l=$\dst #4$}{j}
    \fmfv{d.size=0,l.d=2.0,l.a=-90,l=$\dst #5$}{v2}
  \end{fmfgraph*}\end{center}}}
\newcommand{\ixxf}[5]{\parbox{4cm}{\begin{center}
  \smallfeynmf\begin{fmfgraph*}(25,25)
    \fmftop{i,j}
    \fmfbottom{ij}
    \fmf{#1}{v1,i}
    \fmf{#1}{v1,j}
    \fmf{#2,tension=1.5}{v1,v2}
    \fmf{phantom,tension=2.5}{ij,v2}
    \fmfv{d.size=0,l.d=2.0,l.a=120,l=$\dst #3$}{i}
    \fmfv{d.size=0,l.d=2.0,l.a=60,l=$\dst #4$}{j}
    \fmfv{d.size=0,l.d=2.0,l.a=-90,l=$\dst #5$}{v2}
  \end{fmfgraph*}\end{center}}}
% propagator
\newcommand{\prpl}[4]{\parbox{4cm}{\begin{center}
  \smallfeynmf\begin{fmfgraph*}(25,20)
    \fmfleft{i}
    \fmfright{j}
    \fmf{#1,l.d=5.0,l.s=l,label=$\dst #2$}{i,j}
    \fmfv{d.size=0,l.d=4.0,l.a=180,l=$\dst #3$}{i}
    \fmfv{d.size=0,l.d=4.0,l.a=0,l=$\dst #4$}{j}
  \end{fmfgraph*}\end{center}}}
\newcommand{\prpr}[4]{\parbox{4cm}{\begin{center}
  \smallfeynmf\begin{fmfgraph*}(25,20)
    \fmfleft{j}
    \fmfright{i}
    \fmf{#1,l.d=5.0,l.s=r,label=$\dst #2$}{i,j}
    \fmfv{d.size=0,l.d=4.0,l.a=0,l=$\dst #3$}{i}
    \fmfv{d.size=0,l.d=4.0,l.a=180,l=$\dst #4$}{j}
  \end{fmfgraph*}\end{center}}}
% PS example
\newcommand{\eeqqga}{\parbox{4cm}{\begin{center}
  \smallfeynmf\begin{fmfgraph*}(50,40)
    \fmfleft{,,i,}
    \fmftop{,j,k}
    \fmfbottom{a,b}
    \fmf{fermion}{j,z2,k}
    \fmf{photon,tension=1.5}{z1,z2}
    \fmf{fermion,tension=2.5}{a,g1,z1,b}
    \fmf{gluon}{g1,i}
    \fmfv{d.size=0,l.d=3.0,l.a=180,l=$\dst a$}{a}
    \fmfv{d.size=0,l.d=3.0,l.a=0,l=$\dst b$}{b}
    \fmfv{d.size=0,l.d=3.0,l.a=120,l=$\dst 1$}{i}
    \fmfv{d.size=0,l.d=3.0,l.a=90,l=$\dst 2$}{j}
    \fmfv{d.size=0,l.d=3.0,l.a=60,l=$\dst 3$}{k}
  \end{fmfgraph*}\end{center}}}
\newcommand{\eeqqgb}{\parbox{4cm}{\begin{center}
  \smallfeynmf\begin{fmfgraph*}(50,40)
    \fmfright{,,i,}
    \fmftop{j,k,}
    \fmfbottom{a,b}
    \fmf{fermion}{j,z2,k}
    \fmf{photon,tension=1.5}{z1,z2}
    \fmf{fermion,tension=2.5}{a,z1,g1,b}
    \fmf{gluon}{i,g1}
    \fmfv{d.size=0,l.d=3.0,l.a=180,l=$\dst a$}{a}
    \fmfv{d.size=0,l.d=3.0,l.a=0,l=$\dst b$}{b}
    \fmfv{d.size=0,l.d=3.0,l.a=60,l=$\dst 1$}{i}
    \fmfv{d.size=0,l.d=3.0,l.a=120,l=$\dst 2$}{j}
    \fmfv{d.size=0,l.d=3.0,l.a=90,l=$\dst 3$}{k}
  \end{fmfgraph*}\end{center}}}

%% file: text/abstract.tex
We present a new tree-level matrix element generator,
based on the colour dressed Berends-Giele recursive
relations. We discuss two new algorithms for phase space integration,
dedicated to be used with large multiplicities and colour sampling.

%% file: text/intro.tex
In recent years considerable progress has been made in the calculation
of full matrix elements (ME) for higher order perturbative corrections 
to Standard Model (SM) processes, QCD and QCD associated processes 
in particular. Automatic computation of NLO virtual corrections to arbitrary 
processes finally seems within reach due to newly emerging numerical 
techniques~\cite{Ossola:2006us,*Ellis:2007br,*Giele:2008ve,*Ossola:2008xq,
  *Catani:2008xa,Ossola:2007ax,*Berger:2008sj,*Giele:2008bc}. On-shell recursive 
methods proved to yield compact expressions for multi-leg tree-level amplitudes 
with massless~\cite{Britto:2004ap,*Britto:2005fq} and 
massive~\cite{Badger:2005zh,*Badger:2005jv,*Ozeren:2006ft} 
external particles and are now widely used.
The CSW vertex rules~\cite{Cachazo:2004kj,*Risager:2005vk,*Mansfield:2005yd} 
as off-shell techniques are employed in many analytical and numerical 
approaches~\cite{Badger:2004ty,*Birth:2005vi,*Duhr:2008wc,Gleisberg:2008ft}.

Apart from major developments in the computation of loop amplitudes,
many attempts have been made to tackle the task of numerically evaluating 
tree-level amplitudes with large numbers of external legs. They led
to the construction of several programs, capable of evaluating general
tree-level processes~\cite{Kanaki:2000ey,*Papadopoulos:2000tt,*Cafarella:2007pc,
  Krauss:2001iv,Maltoni:2002qb,*Alwall:2007st,Mangano:2002ea}. In this context 
it turned out, that with increasing number of particles involved in the scattering 
one of the the most efficient methods to compute colour-ordered amplitudes is 
the Berends-Giele recursion~\cite{Berends:1987me,Berends:1987cv,
  *Kleiss:1988ne,*Berends:1988yn,*Berends:1989ie,Dinsdale:2006sq}. 
Correspondingly the fastest methods available for the computation 
of full scattering amplitudes are the colour dressed Berends-Giele 
relations~\cite{Duhr:2006iq}, which are essentially equivalent to the 
Dyson-Schwinger methods employed in Refs.~\cite{Draggiotis:2002hm},
with the {\sc ALPHA} algorithm of Ref.~\cite{Caravaglios:1995cd} being
comparable in efficiency.
In Refs.~\cite{Draggiotis:2002hm} and~\cite{Duhr:2006iq} it was pointed out
that a vertex decomposition of four-gluon vertices in QCD is clearly 
advantageous if the speed of numerical implementations is concerned. 
These findings raise the question,
whether it is possible to construct a full set of SM Feynman rules 
with no four vertices present in the theory, such that recursive
techniques analogous to the colour dressed Berends-Giele relations
can be employed in numerical programs. In Sec.~\ref{sec:sm} 
we demonstrate that this is feasible. We discuss the numerical 
implementation of the results in the new ME generator \Comix in 
Sec.~\ref{sec:me} and present code-related aspects, such as a 
multi-threading concept.

A very important part of computing cross sections for tree-level
processes is, to find an efficient algorithm for phase space generation.
If colours are sampled over, similar problems arise for colour space.
An effective general technique for phase space generation has been presented
in Ref.~\cite{Byckling:1969sx}. 
We observe in Sec.~\ref{sec:recursive_phasespace}, 
that it is possible to formulate the rules presented ibidem in a
truly recursive fashion, i.e.\ on the same footing as the matrix element 
computation. This implies in particular, that point by point the same 
calculational effort is spent for computing matrix element and phase space 
weight. We introduce effective colour sampling techniques 
in Sec.~\ref{sec:colour_sampling}. Having these techniques at hand,
we elaborate on how to eventually couple colour and phase space integration 
and propose a new type of integrator based on the \Haag 
generator~\cite{vanHameren:2002tc} in Sec.~\ref{sec:col_mom_integrator}.

We present a comprehensive comparison of results generated with \Comix 
to those generated with the two other multi-leg tree-level matrix element 
generators \Amegic~\cite{Krauss:2001iv} and \Alpgen~\cite{Mangano:2002ea} 
in Sec.~\ref{sec:results}.
Section~\ref{sec:conclusions} contains our conclusions.

%% file: text/sm.tex
It has been pointed out, for example in 
Refs.~\cite{Draggiotis:2002hm,Dinsdale:2006sq,Duhr:2006iq}, 
that the calculation of multi-parton amplitudes is substantially 
simplified when employing Berends-Giele type recursive relations.
One main reason for the simplification is that these relations allow
to reuse basic building blocks of an amplitude, which are the 
$m$-particle internal off-shell currents. Another reason is that
they can be easily rewritten to include three-particle vertices only. 
In the following we will briefly illuminate, why this is a major advantage.

\mysubsection{The cost of computing a tree amplitude}
\label{sec:computational_cost}
As an example, we try to estimate the total computational cost 
for tree amplitudes, given a certain type of vertices in the underlying theory.
We assume that only one particle type exists and the internal $n$-particle 
currents obey a recursion, which is of the functional form 
\begin{equation}\label{eq:general_recursion}
  J_n\rbr{\pi}=P_n\rbr{\pi}\,\sum\limits_{N=1}^n 
    \sum\limits_{\mathcal{P}_N\rbr{\pi}} V_N\rbr{\pi_1,\ldots,\pi_N}  
    J_{i_1}\rbr{\pi_1}\ldots J_{i_N}\rbr{\pi_N}\;.
\end{equation}
Here $J_m$ denote unordered $m$-particle currents, while $V_N$ are 
$N+1$-point vertices and $P_n$ is a propagator term. The two sums run over 
all possible vertex types $V_N$ and all (unordered) partitions $\mc{P}_N\rbr{\pi}$
of the set of particles $\pi$ into $N$ (unordered) subsets, 
respectively~\cite{Duhr:2006iq}. 
The full $n+1$-particle scattering amplitude can be constructed by putting 
an arbitrary $n$-particle internal off-shell current on-shell and contracting 
the remaining quantity with the corresponding external one-particle current.
\begin{equation}
  A_{n+1}\rbr{\pi}=J_1\rbr{i}\,\frac{1}{P_n\rbr{\pi\setminus i}}\,
    J_n\rbr{\pi\setminus i}\;.
\end{equation}
We now deal only with vertices of $N+1$ external legs and we consider 
their contribution to the computation of an $n$-particle off-shell 
current. The number of vertices to evaluate per $m$-particle subcurrent 
is the Stirling number of the second kind $S\rbr{m,N}$, corresponding to 
the number of partitions of a set $\pi$ of $m$ integers into $N$ subsets. 
The total number $V(n,N)$ of $N+1$-particle vertices to be calculated 
thus becomes
\begin{equation}
  V(n,N)\,=\;\sum\limits_{m=N}^{n}\binom{n}{m}\,S\rbr{m,N}\,.
\end{equation}
Since the Stirling numbers $S(m,N)$ are zero for $m<N$,
we can extend the sum down to zero, leading to
\begin{equation}
  \begin{split}
    V(n,N)\,%=\;\sum\limits_{m=0}^{n}\binom{n}{m}\,S\rbr{m,N}\,
    &=\,\sum\limits_{m=0}^{n}\binom{n}{m}\,
      \frac{1}{N!}\sum\limits_{i=0}^N\rbr{-1}^i\binom{N}{i}\rbr{N-i}^m\\
%    &=\,\frac{1}{N!}\sum\limits_{i=0}^N\rbr{-1}^i\binom{N}{i}\rbr{N+1-i}^n\\
    &=\,\frac{1}{(N+1)!}\sum\limits_{i=0}^N\rbr{-1}^i\binom{N+1}{i}
      \rbr{N+1-i}^{n+1}\,
    =S\rbr{n+1,N+1}\;.
  \end{split}
\end{equation}
The question is, whether we can obtain a milder growth in computational 
complexity, if all $N+1$-particle vertices occuring in 
Eq.~\eqref{eq:general_recursion} are decomposed in terms of two or more vertices 
with fewer number of external legs. When doing so, we must introduce
additional pseudoparticles reflecting the structure of the decomposed
vertex. Hence we have to consider the contribution arising from the
presence of these pseudoparticles, too. 
The problem can be simplified by assuming that there is only one additional 
pseudoparticle, which obeys a completely independent recursion relation.
Then the full contribution of an $N+1$-particle vertex, now being decomposed 
into a $M+1$- and a $N-M+1$-particle vertex becomes
\begin{equation}
  S\rbr{n+1,N+1}\to S\rbr{n+1,M+1}+S\rbr{n+1,N-M+1}\;,
\end{equation}
which can be either bigger or smaller than $S\rbr{n+1,N+1}$, depending on
$n$, $N$ and $M$. With increasing $n$, however the right hand side is always
smaller such that the vertex decomposition becomes clearly advantageous.
Similar arguments hold when introducing more than one pseudoparticle.

From this simple but general consideration we see that the aim of any recursive
formulation of interaction models should be, to reduce the number of external 
lines at interaction vertices to the lowest possible. In this section
we will show that within the Standard Model it is possible to reduce
$N_{\rm max}$ to two, which is the lowest possible number in general.
For QCD interactions we employ the results of Ref.~\cite{Duhr:2006iq},
where this task has already been performed and the original Berends-Giele 
recursive relations have been reformulated to incorporate colour.

\mysubsection{General form of the recursive relations}
\label{sec:general_recursion_me}
In the following we will denote by $\mc{J}_{\alpha}\rbr{\pi}$ 
an unordered SM current of type $\alpha$, which receives contributions
from all Feynman graphs having as external particles the on-shell SM 
particles in the set $\pi$ and one internal particle, described by 
this current. The index $\alpha$ is a multi-index,
carrying information on all quantum numbers and eventually on the
pseudoparticle character of the particle. Special currents are given by 
the external particle currents. They correspond to external scalars,
spinors and polarisation vectors, see Sec.~\ref{sec:me}. For them there
is only one multi-index $\alpha=\alpha_i$ associated with the
external particle $i$, whereas in the general case multiple multi-indices
may lead to non-vanishing internal currents. This corresponds to multiple
particle types being possible as intermediate states. Assuming that only
three-point vertices exist, any internal SM particle and pseudoparticle
off-shell current can be written as
\begin{equation}\label{eq:sm_general_recursion}
  \mc{J}_{\alpha}\rbr{\pi}=P_{\alpha}\rbr{\pi}\,
    \sum\limits_{\mc{V}_{\alpha}^{\;\alpha_1,\,\alpha_2}} 
    \sum\limits_{\mathcal{P}_2\rbr{\pi}}\mc{S}\rbr{\pi_1,\pi_2}\;
    \mc{V}_{\alpha}^{\,\alpha_1,\,\alpha_2}\rbr{\pi_1,\pi_2}\,  
    \mc{J}_{\alpha_1}\rbr{\pi_1}\mc{J}_{\alpha_2}\rbr{\pi_2}\;.
\end{equation}
Here $P_{\alpha}\rbr{\pi}$ denotes a propagator term depending on the
particle type $\alpha$ and the set $\pi$. The term
$\mc{V}_{\alpha}^{\,\alpha_1,\alpha_2}\rbr{\pi_1,\pi_2}$ is a vertex
depending on the particle types $\alpha$, $\alpha_1$ and $\alpha_2$ and 
the decomposition of the set $\pi$ into disjoint subsets $\pi_1$ and $\pi_2$.
The quantity $\mc{S}\rbr{\pi_1,\pi_2}$ is the symmetry factor associated
with the decomposition of $\pi$ into $\pi_1$ and $\pi_2$ and will be
discussed in Sec.~\ref{sec:pref_fermions}.
Superscripts in this context refer to incoming particles, subscripts
to outgoing particles.
The sums run over all vertices in the reformulated Standard Model and
all unordered partitions $\mathcal{P}_2$ of the set $\pi$ into two
disjoint subsets, respectively.
A full unordered $n$-particle scattering amplitude is then given by
\begin{equation}
  \mc{A}\rbr{\pi}=\mc{J}_{\alpha_n}\rbr{n}\,
    \frac{1}{P_{\bar\alpha_n}\rbr{\pi\setminus n}}\,
    \mc{J}_{\bar\alpha_n}\rbr{\pi\setminus n}\;,
\end{equation}
where $\bar\alpha$ denotes a set of reversed particle properties, i.e.\ 
opposite helicity, colour, momentum and particle type.
It has been proved in Ref.~\cite{Duhr:2006iq} that the above form is
correct for pure gluonic scattering amplitudes once the four gluon 
vertex is suitably decomposed into two vertices involving an internal 
antisymmetric tensor pseudoparticle. We briefly recall this proof
before continuing with the decomposition of four particle vertices 
in electroweak interactions. Once this decomposition is achieved,
no further complications arise and Eq.~\eqref{eq:sm_general_recursion}
can be employed to compute arbitrary scattering amplitudes in the
Standard Model.

\mysubsection{Colour dressed Berends-Giele recursive relations in QCD}
Any perturbative QCD scattering amplitude $\mc{A}$ can be written as a sum 
of terms, which factorise into two components, one only depending on the gauge
structure and one only depending on the kinematics. Such a decomposition
is called colour decomposition. Considering for example tree-level $n$-gluon 
amplitudes, several colour decompositions exist. A very intuitive one based 
on the fundamental representation of the gauge group is given 
by~\cite{Mangano:1987xk} 
\begin{equation}\label{eq:color_decomposition_fundamental}
  \mc{A}\rbr{1,\ldots,n}=\sum\limits_{\vec\sigma\in S_{n-1}}
    {\rm Tr}\rbr{T^{a_1}T^{a_{\sigma_2}}\ldots T^{a_{\sigma_n}}}\;
    A\rbr{1,\sigma_2,\ldots,\sigma_n}\;.
\end{equation}
Here $\vec\sigma$ runs over all permutations $S_{n-1}$ of the $n-1$ indices
$2\ldots n$.
The functions $A$ depend on the Lorentz-structure of the process only and
are called colour-ordered amplitudes. A more suitable colour decomposition
for $n$-gluon amplitudes has been introduced in 
Refs.~\cite{DelDuca:1999rs,*DelDuca:1999ha}. It employs the adjoint
representation matrices $(F^a)_{bc}$ of $SU(3)$ and reads
\begin{equation}\label{eq:color_decomposition_adjoint}
  \mc{A}\rbr{1,\ldots,n}=\sum\limits_{\vec\sigma\in S_{n-2}}
    \rbr{F^{a_{\sigma_2}}\ldots F^{a_{\sigma_{n-1}}}}_{a_1a_n}\;
    A\rbr{1,\sigma_2,\ldots,\sigma_{n-1},n}\;.
\end{equation}
Note that in this case the sum runs over the permutations of the $n-2$
indices $2\ldots n-1$ only, whereas the first and the last index remain 
fixed. Another colour decomposition, suited especially for Monte Carlo 
event generation is the colour flow decomposition~\cite{Maltoni:2002mq}.
In this prescription the $SU(3)$ gluon field is treated as a $3\times 3$
matrix $(A_\mu)^{i\bar{\jmath}}$ rather than a one index field $A_\mu^a$. 
The corresponding decomposition reads
\begin{equation}\label{eq:color_decomposition_colorflow}
  \mc{A}\rbr{1,\ldots,n}=\sum\limits_{\vec\sigma\in S_{n-1}}
    \delta^{i_1\bar{\jmath}_{\sigma_2}}
    \delta^{i_{\sigma_2}\bar{\jmath}_{\sigma_3}}
    \ldots\delta^{i_{\sigma_n}\bar{\jmath}_1}\;
    A\rbr{1,\sigma_2,\ldots,\sigma_n}\;.
\end{equation}
The remaining task is now, to compute the colour-ordered amplitudes.
In Ref.~\cite{Berends:1987me} Berends and Giele proposed a method to do so
in a recursive fashion. The basic idea is that, according to the Feynman 
rules of QCD, an internal $n$-gluon current is defined by all contributing 
Feynman graphs with $n$ external on-shell gluons and one off-shell gluon.
\begin{equation}\label{eq:cobg}
  \begin{split}
    &J_\mu\rbr{1,2,\ldots,n}=\frac{-ig_{\mu\nu}}{p_{1,n}^2}\left\{\;
      \vphantom{\sum_{k=j+1}^{n-1}}
      \sum_{k=1}^{n-1}V_3^{\nu\kappa\lambda}\rbr{p_{1,k},p_{k+1,n}}
        J_\kappa\rbr{1,\ldots,k}J_\lambda\rbr{k+1,\ldots,n}\right.\\
      &\qquad\qquad+\left.\sum_{j=1}^{n-2}\sum_{k=j+1}^{n-1}
        V_4^{\nu\rho\kappa\lambda}J_\rho\rbr{1,\ldots,j}
        J_\kappa\rbr{j+1,\ldots,k}J_\lambda\rbr{k+1,\ldots,n}\right\}\;.
  \end{split}
\end{equation}
Here $p_i$ denote the momenta of the gluons, $p_{i,j}=p_i+\ldots+p_j$ and 
$V_3^{\nu\kappa\lambda}\left(p_{1,k},p_{k+1,n}\right)$ and 
$V_4^{\nu\rho\kappa\lambda}$ are the colour-ordered three and four-gluon 
vertices defined according to Ref.~\cite{Dixon:1996wi}, 
\begin{equation}
  \begin{split}
    V_3^{\nu\kappa\lambda}\rbr{p,q}&=i\,\frac{g_s}{\sqrt{2}}
      \rbr{\,g^{\kappa\lambda}\rbr{p-q}^\nu+
        g^{\lambda\nu}\rbr{2q+p}^\kappa-g^{\nu\kappa}\rbr{2p+q}^\lambda\,}\;,\\
    V_4^{\nu\rho\kappa\lambda}&=i\,\frac{g_s^2}{2}
      \rbr{\,2g^{\nu\kappa}g^{\rho\lambda}-
        g^{\nu\rho}g^{\kappa\lambda}-g^{\nu\lambda}g^{\rho\kappa}\,}\;.
  \end{split}
\end{equation}
The full colour-ordered $n$-gluon amplitude $A\rbr{1,\ldots,n}$ is then 
obtained by putting the $n-1$-particle off-shell current 
$J_{n-1}\rbr{1,\ldots,n-1}$ on-shell and contracting it with the external 
polarisation $J_\mu\rbr{n}$.
Employing the tensor-gluon vertex
\begin{equation}
  V_T^{\mu\nu\kappa\lambda}=\frac{i}{2}\frac{g_s}{\sqrt{2}}
    \rbr{g^{\mu\kappa}g^{\nu\lambda}-g^{\mu\lambda}g^{\nu\kappa}}\;,
\end{equation}
and the tensor ``propagator''
\begin{equation}\label{eq:prop_pseudo}
  -iD_{\mu\nu}^{\,\kappa\lambda}=
    -i\rbr{g_\mu^\kappa g_\nu^\lambda-g_\mu^\lambda g_\nu^\kappa}\;,
\end{equation}
the recursion can be reformulated to give
\begin{equation}
  \begin{split}
    &J_\mu\rbr{1,2,\ldots,n}=\frac{-ig_{\mu\nu}}{p_{1,n}^2}
      \sum_{k=1}^{n-1}\left\{\vphantom{\sum_{k=1}^{n-1}}\;
        V_3^{\nu\kappa\lambda}\rbr{p_{1,k},p_{k+1,n}}
        J_\kappa\rbr{1,\ldots,k}J_\lambda\rbr{k+1,\ldots,n}\right.\\
      &\qquad\qquad+V_T^{\nu\kappa\alpha\beta}
        J_\kappa\rbr{1,\ldots,k}J_{\alpha\beta}\rbr{k+1,\ldots,n}
      +\left.V_T^{\lambda\nu\alpha\beta}J_{\alpha\beta}\rbr{1,\ldots,k}
        J_\lambda\rbr{k+1,\ldots,n}\vphantom{\sum_{k=1}^{n-1}}\right\}
  \end{split}
\end{equation}
and
\begin{equation}
  J^{\alpha\beta}\rbr{1,2,\ldots,n}=
    -iD^{\,\alpha\beta}_{\gamma\delta}\,
    \sum_{k=1}^{n-1}V_T^{\gamma\delta\kappa\lambda}
      J_\kappa\rbr{1,\ldots,k}J_\lambda\rbr{k+1,\ldots,n}\;,
\end{equation}
for the gluon and tensor pseudoparticle currents, respectively.
Since no external tensor currents exist, all tensor currents with
one particle index only are defined as zero. The advantage of the 
above formulation including a tensor current, as discussed in
Sec.~\ref{sec:computational_cost}, is the elimination of the 
four-gluon vertex. Correspondingly we introduce a ``pseudogluon'',
which, from here on, we denote by $g_4$.

Following Ref.~\cite{Duhr:2006iq}, one can introduce colour dressed
gluon and tensor pseudoparticle currents $\mc{J}_{\mu\,I\bar{J}}$ and 
$\mc{J}_{\alpha\beta\,I\bar{J}}$, defined by
\begin{equation}
  \begin{split}
    \mc{J}_{\mu\,I\bar{J}}\rbr{1,\ldots,n}&=
    \sum\limits_{\vec\sigma\in S_n}\delta_{I\bar{\jmath}_{\sigma_1}}
    \delta_{i_{\sigma_1}\bar{\jmath}_{\sigma_2}}
    \ldots\delta_{i_{\sigma_n}\bar{J}}\;
    J_\mu\rbr{\sigma_1,\ldots,\sigma_n}\;,\\
    \mc{J}_{\alpha\beta\,I\bar{J}}\rbr{1,\ldots,n}&=
    \sum\limits_{\vec\sigma\in S_n}\delta_{I\bar{\jmath}_{\sigma_1}}
    \delta_{i_{\sigma_1}\bar{\jmath}_{\sigma_2}}
    \ldots\delta_{i_{\sigma_n}\bar{J}}\;
    J_{\alpha\beta}\rbr{\sigma_1,\ldots,\sigma_n}\;.
  \end{split}
\end{equation}
Denoting by $\pi$ the set $\rbr{1,\ldots,n}$ of $n$ particles,
the following recursive relations for these currents are obtained:
\begin{equation}
  \begin{split}
    \mc{J}_{\mu\,I\bar J}\rbr{\pi}&=
      D_{\mu\,I\bar J}^{\,\nu\,H\bar G}\rbr{\pi}\left\{\;
      \sum_{\mc{P}_2\rbr{\pi}}
        \mc{V}_{\nu\,H\bar G}^{\,
	  \kappa\,K\bar L,\,\lambda\,M\bar N}\rbr{\pi_1,\pi_2}\,
        \mc{J}_{\kappa\,K\bar L}\rbr{\pi_1}
        \mc{J}_{\lambda\,M\bar N}\rbr{\pi_2}\right.\\
      &\text{\hspace*{18ex}}+\left.\sum_{\mc{OP}_2\rbr{\pi}}
        \mc{V}_{\nu\, H\bar G}^{\,\kappa\,K\bar L,\,\alpha\beta\,M\bar N}\,
        \mc{J}_{\kappa\,K\bar L}\rbr{\pi_1}
        \mc{J}_{\alpha\beta\,M\bar N}\rbr{\pi_2}\;\right\}\;,\\
  \mc{J}_{\alpha\beta\,I\bar J}\rbr{\pi}&=
    D_{\alpha\beta\,I\bar J}^{\,\gamma\delta\,H\bar G}\,
    \sum_{\mc{P}_2\rbr{\pi}}
    \mc{V}_{\gamma\delta\,H\bar G}^{\,\kappa\,K\bar L,\,\lambda\,M\bar N}\,
    \mc{J}_{\kappa\,K\bar L}\rbr{\pi_1}\mc{J}_{\lambda\,M\bar N}\rbr{\pi_2}\;.
  \end{split}
\end{equation}
Here we have defined the colour dressed gluon and tensor pseudoparticle
vertices
\begin{equation}
  \begin{split}
    \mc{V}_{\nu\,H\bar G}^{\,\kappa\,K\bar L,\,\lambda\,M\bar N}\rbr{\pi_1,\pi_2}=
      \delta^{\bar L}_{\bar G}\delta^{K\bar N}\delta_H^M\,
      V_{3\,\nu}^{\;\;\,\kappa\lambda}\rbr{\pi_1,\pi_2}+
      \delta_H^K\delta^{M \bar L}\delta^{\bar N}_{\bar G}\,
      V_{3\,\nu}^{\;\;\,\lambda\kappa}\rbr{\pi_2,\pi_1}\;,
  \end{split}
\end{equation}
and
\begin{equation}
  \begin{split}
    \mc{V}_{\gamma\delta\,H\bar G}^{\,\kappa\,K\bar L,\,\lambda\,M\bar N}=
      \delta^{\bar L}_{\bar G}\delta^{K\bar N}\delta_H^M\,
      V_{T\,\gamma\delta}^{\;\;\,\kappa\lambda}+
      \delta_H^K\delta^{M \bar L}\delta^{\bar N}_{\bar G}\,
      V_{T\,\gamma\delta}^{\;\;\,\lambda\kappa}\;.
  \end{split}
\end{equation}
The second sum runs over the set of ordered partitions of the set $\pi$
into two disjoint subsets, $\mc{OP}_2(\pi)$.
A complete proof of these relations can be found in Ref.~\cite{Duhr:2006iq}.
The above procedure of colour dressing can easily be generalised to QCD 
processes including quarks. Since no further elementary QCD four-point 
interactions exists, no further vertex decomposition has to be performed
and therefore no new current types are introduced. For amplitudes including
quarks care must be taken of using the proper colour space gluon propagator 
when coupling to $q\bar{q}g$ vertices, i.e.
\begin{equation}\label{eq:gluon_colorprop_full}
  P_{g\, I\bar J}^{\hphantom{g}\,H\bar G}\;\varpropto\;
    \delta_I^H\delta_{\bar J}^{\bar G}
    -\frac{1}{N_C}\delta_{I\bar J}\,\delta^{H\bar G}\;,
\end{equation}
as described in Ref.~\cite{Maltoni:2002mq}.

\mysubsection{Decomposition of electroweak four-particle vertices}
The above procedure can be generalised to describe all Standard Model 
interactions, once a suitable replacement of the corresponding four particle
vertices has been found.

We start by proposing a decomposition of four particle vertices with 
$W$-bosons only\footnote{ Note that this decomposition of vertices
  is not unique and other choices may exist.}
\begin{equation}
  \begin{split}
    \mc{V}_{W^-\mu}^{\,W^-\kappa,\,W^+\nu,\,W^-\lambda}\to&\;
      \mc{V}_{W^-\mu}^{\,W^-\kappa,\,Z_4\gamma\delta}
      \cdot P_{Z_4\,\gamma\delta}^{\hphantom{Z_4}\,\alpha\beta}
      \cdot\mc{V}_{Z_4\alpha\beta}^{\,W^+\nu,\,W^-\lambda}\,+\;
      \mc{V}_{W^-\mu}^{\,W^-\lambda,\,Z_4\gamma\delta}
      \cdot P_{Z_4\,\gamma\delta}^{\hphantom{Z_4}\,\alpha\beta}
      \cdot\mc{V}_{Z_4\alpha\beta}^{\,W^+\nu,\,W^-\kappa}\;.
  \end{split}
\end{equation}
Here $Z_4$ denotes a new antisymmetric tensor pseudoparticle introduced 
for the vertex decomposition. Its interaction vertex reads
\begin{equation}
  \begin{split}
    \mc{V}_{W^-\mu}^{\,W^-\kappa,\,Z_4\gamma\delta}&=\frac{i}{2}\,g_w\,
      \rbr{g_\mu^\gamma g^{\kappa\delta}-g_\mu^\delta g^{\kappa\gamma}}\;,&
    \mc{V}_{Z_4\alpha\beta}^{\,W^+\kappa,\,W^-\lambda}&=\frac{i}{2}\,g_w\,
      \rbr{g_\alpha^\kappa g_\beta^\lambda-g_\alpha^\lambda g_\beta^\kappa}\;.
  \end{split}
\end{equation}
To obtain correct signs of four-particle vertices, we define the
tensor pseudoparticle ``propagators'' as
\begin{equation}
  P_{\alpha\,\mu\nu}^{\hphantom{\alpha}\,\kappa\lambda}=
    \kappa_\alpha D_{\mu\nu}^{\,\kappa\lambda}
  \quad\quad{\rm where}\quad\quad
  \kappa_\alpha=\left\{\begin{array}{cc}-i & {\rm if}\quad\alpha\,=\,Z_4\\
    i & {\rm else}\end{array}\right.\;,
\end{equation}
and where $D_{\mu\nu}^{\,\kappa\lambda}$ is given by Eq.~\eqref{eq:prop_pseudo}.
Note that the $Z_4$ pseudoparticle is not self-conjugate. This 
definition prevents double counting four-particle vertices involving 
the $W$ boson and constructing fake $WWWW$ vertices with all $W$'s having
the same charge. The four-particle vertices involving $W$ bosons, 
photons and $Z$-bosons are decomposed as follows
\begin{equation}
  \begin{split}
    \mc{V}_{W^-\mu}^{\,A\kappa,\,W^-\nu,\,A\lambda}\to&\;
      \mc{V}_{W^-\mu}^{\,A\kappa,\,W_4^-\gamma\delta}
      \cdot P_{W_4^-\,\gamma\delta}^{\hphantom{W_4^-}\,\alpha\beta}
      \cdot\mc{V}_{W_4^-\alpha\beta}^{\,W^-\nu,\,A\lambda}\,+\;
      \mc{V}_{W^-\mu}^{\,A\lambda,\,W_4^-\gamma\delta}
      \cdot P_{W_4^-\,\gamma\delta}^{\hphantom{W_4^-}\,\alpha\beta}
      \cdot\mc{V}_{W_4^-\alpha\beta}^{\,W^-\nu,\,A\kappa}\;,\\
    \mc{V}_{W^-\mu}^{\,A\kappa,\,W^-\nu,\,Z\lambda}\to&\;
      \mc{V}_{W^-\mu}^{\,A\kappa,\,W_4^-\gamma\delta}
      \cdot P_{W_4^-\,\gamma\delta}^{\hphantom{W_4^-}\,\alpha\beta}
      \cdot\mc{V}_{W_4^-\alpha\beta}^{\,W^-\nu,\,Z\lambda}\,+\;
      \mc{V}_{W^-\mu}^{\,Z\lambda,\,W_4^-\gamma\delta}
      \cdot P_{W_4^-\,\gamma\delta}^{\hphantom{W_4^-}\,\alpha\beta}
      \cdot\mc{V}_{W_4^-\alpha\beta}^{\,W^-\nu,\,A\kappa}\;,\\
    \mc{V}_{W^-\mu}^{\,Z\kappa,\,W^-\nu,\,Z\lambda}\to&\;
      \mc{V}_{W^-\mu}^{\,Z\kappa,\,W_4^-\gamma\delta}
      \cdot P_{W_4^-\,\gamma\delta}^{\hphantom{W_4^-}\,\alpha\beta}
      \cdot\mc{V}_{W_4^-\alpha\beta}^{\,W^-\nu,\,Z\lambda}\,+\;
      \mc{V}_{W^-\mu}^{\,Z\lambda,\,W_4^-\gamma\delta}
      \cdot P_{W_4^-\,\gamma\delta}^{\hphantom{W_4^-}\,\alpha\beta}
      \cdot\mc{V}_{W_4^-\alpha\beta}^{\,W^-\nu,\,Z\kappa}\;.
  \end{split}
\end{equation}
We introduced a new tensor pseudoparticle, $W^-_4$, whose interaction 
vertices are defined as
\begin{equation}
  \begin{split}
    \mc{V}_{W^-\mu}^{\,A\kappa,\,W_4^-\gamma\delta}&=
      \frac{i}{2}\,g_w\sin\theta_W\,
      \rbr{g_\mu^\gamma g^{\kappa\delta}-g_\mu^\delta g^{\kappa\gamma}}\;,&
    \mc{V}_{W_4^-\alpha\beta}^{\,W^-\nu,\,A\kappa}&=
      \frac{i}{2}\,g_w\sin\theta_W\,
      \rbr{g_\alpha^\nu g_\beta^\kappa-g_\alpha^\kappa g_\beta^\nu}\;,\\
    \mc{V}_{W^-\mu}^{\,Z\kappa,\,W_4^-\gamma\delta}&=
      \frac{i}{2}\,g_w\cos\theta_W\,
      \rbr{g_\mu^\gamma g^{\kappa\delta}-g_\mu^\delta g^{\kappa\gamma}}\;,&
    \mc{V}_{W_4^-\alpha\beta}^{\,W^-\nu,\,Z\kappa}&=
      \frac{i}{2}\,g_w\cos\theta_W\,
      \rbr{g_\alpha^\nu g_\beta^\kappa-g_\alpha^\kappa g_\beta^\nu}\;.
  \end{split}
\end{equation}
Corresponding vertices exist for $W^+$ / $W^-$ bosons.
The decomposition of four particle vertices involving the Higgs boson
introduces a new scalar pseudoparticle, which we denote by $h_4$. 
In order not to generate fake four particle vertices we define it not
to be self-conjugate. The corresponding vertices read
\begin{equation}
  \begin{split}
    \mc{V}_{h}^{\,h,\,h,\,h}&\to\mc{V}_{h}^{\,h,\,h_4}
      \cdot P_{h_4}\cdot\mc{V}_{h_4}^{\,h,\,h}\;,\\
    \mc{V}_{h}^{\,h,\,Z\mu,\,Z\nu}&\to\mc{V}_{h}^{\,h,\,h_4}
      \cdot P_{h_4}\cdot\mc{V}_{h_4}^{\,Z\mu,\,Z\nu}\;,\\
    \mc{V}_{h}^{\,h,\,W^+\mu,\,W^-\nu}&\to\mc{V}_{h}^{\,h,\,h_4}
      \cdot P_{h_4}\cdot\mc{V}_{h_4}^{\,W^+\mu,\,W^-\nu}\;.
  \end{split}
\end{equation}
where the interactions of the $h_4$ pseudoparticle are defined by
\begin{equation}
  \begin{split}
    \mc{V}_{h}^{\,h,\,h_4}&=i\;,\\
    \mc{V}_{h_4}^{\,h,\,h}&=i\,\frac{m_h^2}{v^2}\;,\\
    \mc{V}_{h_4}^{\,Z\mu,\,Z\nu}&=
      -i\,\frac{g_w^2}{2\,\cos^2\theta_W}\,g^{\mu\nu}\;,\\
    \mc{V}_{h_4}^{\,W^+\mu,\,W^-\nu}&=
      -i\,\frac{g_w^2}{2}\,g^{\mu\nu}\;,\\
  \end{split}
\end{equation}
and where we have introduced the scalar ``propagator'' of the 
$h_4$ pseudoparticle
\begin{equation}
  P_{h_4}=i\;.
\end{equation}
Since all remaining vertices in the Standard Model are three point 
vertices, the vertex decomposition is hereby complete. The additional 
Standard Model propagators and vertices arising from this decomposition
are summarised in Tabs.~\ref{tab:addprops} and~\ref{tab:addvtcs}, 
respectively.
\mytable{t}{
\begin{tabular}{c@{\hspace{3ex}=\hspace{3ex}}l@{\hspace{2ex}}
                c@{=\hspace{3ex}}l@{\hspace{2ex}}}
  \prpr{dots}{g_4}{\kappa\lambda,,H\bar G}{\mu\nu,,I\bar J} & 
  $\dst -i\,\delta_I^H\delta_{\bar J}^{\bar G}\,
    D_{\mu\nu}^{\,\kappa\lambda}$ &
  \prpr{dots}{Z_4}{\kappa\lambda}{\mu\nu} & 
  $\dst -i\,D_{\mu\nu}^{\,\kappa\lambda}$ \\
  \prpr{dashes}{h_4}{}{} & 
  $\dst i$ &
  \prpr{dots}{W_4^\pm}{\kappa\lambda}{\mu\nu} &
  $\dst i\,D_{\mu\nu}^{\,\kappa\lambda}$ \\
\end{tabular}}{
Standard Model propagators for auxiliary particles introduced
in the vertex decomposition. Note that the $1/N_C$-term arising
in Eq.~\eqref{eq:gluon_colorprop_full} is obsolete for the $g_4$ 
pseudogluon propagator because the pseudogluon does not couple 
to quarks.\label{tab:addprops}}
\mytable{t}{
\begin{tabular}{c@{\hspace{1ex}=\hspace{3ex}}l@{\hspace{2ex}}}
  \ixxf{gluon}{dots}{g,,\varepsilon,,K\bar L}{
    g,,\varepsilon',,M\bar N}{g_4,,H\bar G} & 
  $\dst i\,\frac{g_s}{\sqrt{2}}\,\sbr{\;
     \delta_H^K\delta^{\bar L M}\delta_{\bar G}^{\bar N}
     -\delta_H^M\delta^{K\bar N}\delta_{\bar G}^{\bar L}\;}\;
     {\rm VVT}\rbr{\varepsilon,\varepsilon'}$ \\
  \ixxs{dashes}{dashes}{h}{h}{h_4} & 
  $\dst i\,\frac{m_h^2}{v^2}$ \hspace*{1cm}
  \ixxs{dashes}{dashes}{h}{h_4}{h}\hspace*{1ex}=\hspace*{3ex}
  $\dst i$ \\
  \ixxs{wiggly}{dashes}{W/Z,,\varepsilon}{W/Z,,\varepsilon'}{h_4} & 
  $\dst -i\,\frac{g_w^2}{2\,\lambda_{W/Z}^2}\,
    {\rm VVS}\rbr{\varepsilon,\varepsilon'}
    \quad\quad{\rm where}\quad\quad\begin{array}{rc}\lambda_W\,=&1\\
      \lambda_Z\,=&\cos\theta_W\end{array}$ \\
  \ixxs{wiggly}{dots}{W^-,,\varepsilon}{W^+,,\varepsilon'}{Z_4} & 
  $\dst i\,g_w\,{\rm VVT}\rbr{\varepsilon,\varepsilon'}$ \\
  \ixxs{wiggly}{dots}{W^-,,\varepsilon}{A/Z,,\varepsilon'}{W_4^-} & 
  $\dst i\,g_w\,\kappa_{A/Z}\,{\rm VVT}\rbr{\varepsilon,\varepsilon'}
    \quad\quad{\rm where}\quad\quad
    \begin{array}{rl}\kappa_A&=\,\sin\theta_W\\
      \kappa_Z&=\,\cos\theta_W\end{array}$ \\[15mm]
  \multicolumn{1}{c}{${\rm VVS}\rbr{\varepsilon,\varepsilon'}=
  \varepsilon^\mu\varepsilon'_\mu$ ,} &
  ${\rm VVT}^{\,\mu\nu}\rbr{\varepsilon,\varepsilon'}
  =\frac{1}{2}\rbr{g^{\mu\lambda}g^{\nu\kappa}-g^{\mu\kappa}g^{\nu\lambda}}
  \varepsilon_\lambda\varepsilon'_\kappa$\\[5mm]
\end{tabular}\vspace*{2ex}}{
Standard Model vertices arising from the vertex decomposition and
replacing the four particle vertices. In this context $\varepsilon$ and
$\varepsilon'$ denote arbitrary incoming vector currents. 
Note that due to the antisymmetry of ${\rm VVT}^{\,\mu\nu}$, we can make 
the replacement $D_{\alpha\beta}^{\mu\nu}\,{\rm VVT}^{\,\alpha\beta}=
  2\,{\rm VVT}^{\,\mu\nu}$, which leads to a slight decrease 
in computation time.
\label{tab:addvtcs}}

\mysubsection{Prefactors of diagrams with external fermions}
\label{sec:pref_fermions}
When calculating currents with an arbitrary number of possibly 
indistinguishable external fermions, we have to take into account, 
that each Feynman diagram contains a prefactor 
\begin{equation}\label{eq:fermion_sign}
  \mc{S}=(-1)^{\;{\rm S}_f\rbr{\sigma_1,\ldots,\sigma_n}}\;,
\end{equation}
according to the number of fermion permutations ${\rm S}_f$ in the external 
particle assignment $\vec\sigma=\rbr{\sigma_1,\ldots,\sigma_n}$.
To be used in the context of a recursive computation, this prefactor 
must be defined on a local basis in order to avoid the proliferation 
of information on different $\vec\sigma$. It is then sufficient 
to note that Eq.~\eqref{eq:fermion_sign} holds at the
level of interaction vertices. More precisely we can define
the local prefactor $\mc{S}\rbr{\pi_1,\pi_2}$ of 
Eq.~\eqref{eq:sm_general_recursion} as
\begin{equation}\label{eq:fermion_sign_vtx}
  \mc{S}\rbr{\pi_1,\pi_2}=(-1)^{\;{\rm S}_f\rbr{\pi_1,\pi_2}}\;.
\end{equation}
Here ${\rm S}_f\rbr{\pi_1,\pi_2}$ counts the number of fermion permutations 
that is needed to restore a predefined, for example ascending index 
ordering when combining the sets $\pi_1$ and $\pi_2$ into the set 
$\pi=\pi_1\oplus\pi_2$. Upon iterating this procedure, we obtain 
the correct relative prefactors $\mc{S}$ for each diagram.

%% file: text/me.tex
The general formulae to recursively compute tree-level amplitudes have
been stated in Sec.~\ref{sec:sm}. Here we briefly explain, which 
conventions are used to define the external particle currents and internal 
Lorentz structures. We also elaborate on how to organise the computation
and how to reduce the effective computation time per phase space point 
by a multi-threaded structure of the implementation.

\mysubsection{Choice of the spinor basis}
We employ the spinor basis introduced in Ref.~\cite{Hagiwara:1985yu}.
Accordingly, the $\gamma$-matrices are taken in the Weyl representation. 
The main advantage of this representation is that spinors for massless 
particles are described through two nonzero components only. This fact
greatly alleviates their construction as well as the evaluation of
vertices. Polarisation vectors for external vector bosons are constructed 
according to Ref.~\cite{Dittmaier:1998nn}.
As pointed out in Sec.~\ref{sec:sm}, within the Standard Model tensor particles 
never occur as external states, such that there is no need to explicitly 
construct polarisation tensors.

\mysubsection{Implementation details}
\label{sec:implementation_me}
The algorithms presented in this paper are intended to be used for large
multiplicity matrix element calculations. In this context, it is often useful
to sample over helicities of external particles in a Monte Carlo fashion.
However, this introduces additional degrees of freedom and leads to a slower
convergence of the integral. Furthermore when taking 
Eq.~\eqref{eq:sm_general_recursion} serious, we note that for  
helicity-summed ME's, it is possible to reuse currents to
compute amplitudes with different configurations. Namely if the 
helicities of external particles assigned to a particular current do not 
change, it does not need to be recomputed.
This leads to a significant decrease in evaluation time for the helicity
summed ME's compared to the naive method of computing the full
amplitude afresh for different configurations. A corresponding comparison
can be found in Sec.~\ref{sec:results}. The default choice in \Comix is
helicity summation. To allow computations for very large multiplicities,
however, helicity sampling can be enabled as an option. 
\myfigure{t}{\includegraphics[width=12cm]{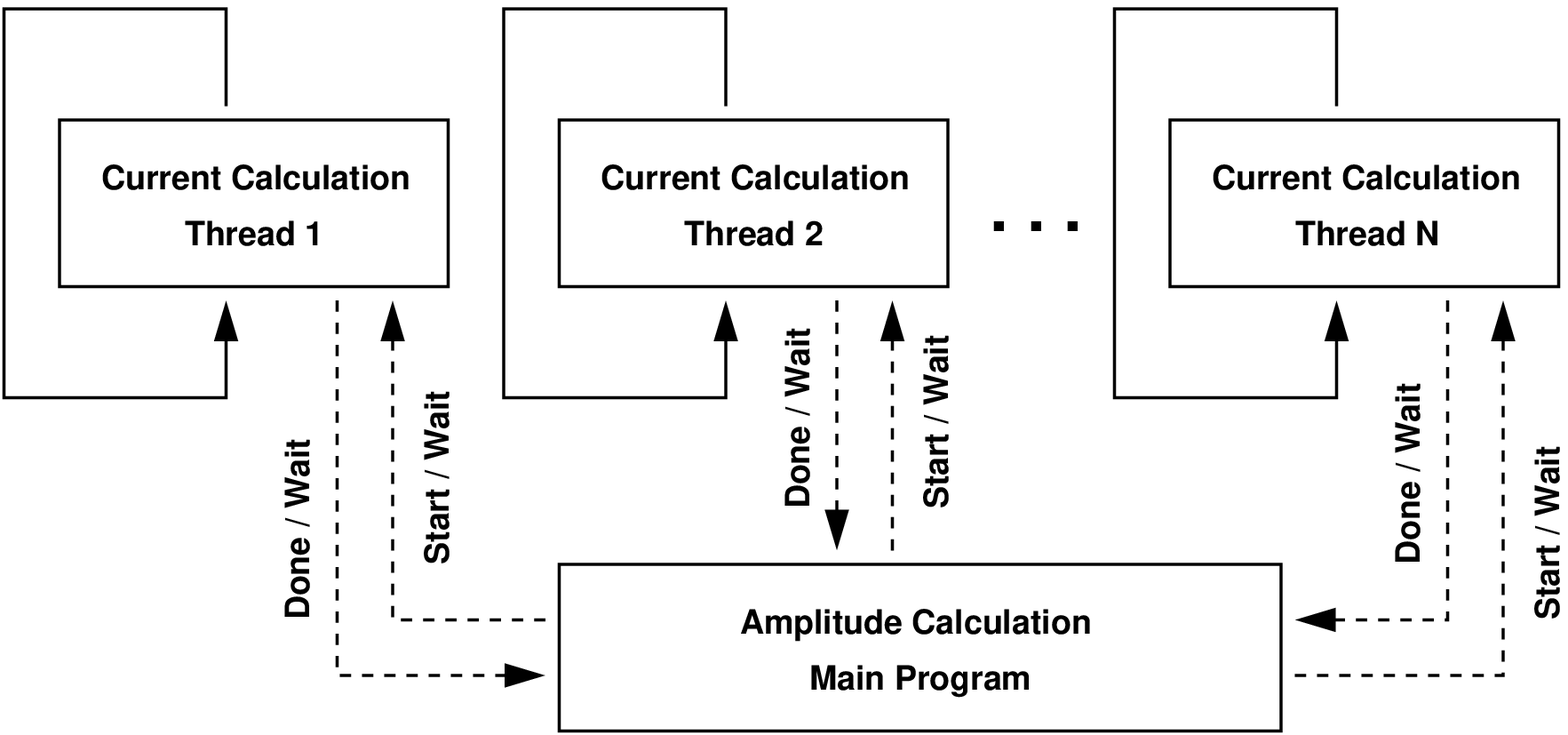}
  \vspace*{5mm}}{Structure of the multi-threaded implementation 
  for matrix element computation in \Comix. The number of threads $N$ 
  is variable and depends on the number of available processors. 
  The main program communicates start and wait signals to the 
  calculator threads, while those communicate done and wait 
  signals to the main program. Details are explained  in the text.
  \label{fig:multithread_me}}

The effective computation time per phase space point
can be further reduced by a multi-threaded implementation of 
Eq.~\eqref{eq:sm_general_recursion}. Figure~\ref{fig:multithread_me} shows 
the basic structure of this algorithm. The main advantage of 
Eq.~\eqref{eq:sm_general_recursion} is, that in order to compute a current
that depends on $n$ external particles, it is sufficient to know all
subcurrents that depend on $m<n$ external particles. This leads to a
straightforward multi-threading algorithm.
\begin{itemize}
  \item Create N threads at program startup with the following properties
  \begin{enumerate}
    \item\label{thread_entrypoint}
      The thread waits for the main program to signal the start 
      of a computation.\\ It then signals the main program to wait.
    \item It takes a number $n$ and computes a block of currents 
      depending on $n$ external particles using subcurrents depending
      on $m<n$ external particles. If $n=1$, it computes external
      polarisation vectors and spinors.
    \item It signals the main program that the calculation is done
      and returns to step~\ref{thread_entrypoint}.
  \end{enumerate}
  \item For each phase space point, employ the following algorithm
    in the main program
  \begin{enumerate}
    \item Start with $n=1$.
    \item\label{main_entrypoint} 
      Split the number of currents that depend on $n$ external particles
      into $N$ blocks.\\ Communicate $n$ and one block to each calculator 
      thread.
    \item Signal the threads to start their computation. \\
      Wait for all threads to signal completion.
    \item Let $n\to n+1$ and return to step~\ref{main_entrypoint} 
      if further currents need to be computed.
  \end{enumerate}
\end{itemize}
The efficiency of this algorithm solely depends on an efficient
thread library. The overhead with a modern POSIX threading
is about 10\% of the total computational cost. 
This, however, is not of any concern considering that the employment of multiple CPU's
reduces the computation time roughly proportional to the increase in processor usage.

%% file: text/ps.tex
\newcommand{\insertgenvertices}{
\myfigure{tb}{
  \begin{picture}(140,121) (45,-45)
    \SetWidth{0.5}
    \SetColor{Black}
    \ArrowLine(120,26)(165,61)
    \ArrowLine(120,26)(75,61)
    \ArrowLine(120,-29)(120,26)
    \GOval(165,61)(13,13)(0){0.882}
    \GOval(75,61)(13,13)(0){0.882}
    \COval(120,26)(19,19)(45.0){Black}{White}
    \GOval(120,-29)(13,13)(0){0.666}
    \Text(75,61)[]{\Black{$\rho$}}
    \Text(165,61)[]{\Black{$\pi\setminus\rho$}}
    \Text(120,-29)[]{\Black{$\pi$}}
    \Text(120,26)[]{\Black{$\bar{S}_{\pi}^{\,\rho,\pi\setminus\rho}$}}
  \end{picture}
  \begin{picture}(140,121) (45,-45)
    \SetWidth{0.5}
    \SetColor{Black}
    \DashArrowLine(120,16)(165,61){5}
    \ArrowLine(120,16)(75,61)
    \ArrowLine(75,-29)(120,16)
    \ArrowLine(165,-29)(120,16)
    \GOval(75,61)(13,13)(0){0.888}
    \GOval(165,61)(13,13)(0){0.888}
    \COval(120,16)(19,19)(45.0){Black}{White}
    \Text(75,61)[]{\Black{$\pi$}}
    \Text(165,61)[]{\Black{$\overline{\alpha b\pi}$}}
    \GOval(75,-29)(13,13)(0){0.666}
    \Text(165,-29)[lt]{\Black{$\hspace*{1ex}b$}}
    \Text(75,-29)[]{\Black{$\alpha$}}
    \Text(120,16)[]{\Black{$\bar{T}_{\alpha,b}^{\,\pi,\overline{\alpha b\pi}}$}}
  \end{picture}
  \begin{picture}(140,121) (45,-45)
    \SetWidth{0.5}
    \SetColor{Black}
    \ArrowLine(120,10)(120,61)
    \ArrowLine(75,-29)(120,6)
    \ArrowLine(165,-29)(120,6)
    \GOval(120,61)(13,13)(0){0.888}
    \COval(120,6)(19,19)(45.0){Black}{White}
    \Text(120,61)[]{\Black{$\overline{\alpha b}$}}
    \GOval(75,-29)(13,13)(0){0.666}
    \Text(165,-29)[lt]{\Black{$\hspace*{1ex}b$}}
    \Text(75,-29)[]{\Black{$\alpha$}}
    \Text(120,6)[]{\Black{$\bar{D}_{\alpha,b}$}}
  \end{picture}}{
  Basic vertices for phase space generation. Grey blobs
  correspond to eventually off mass-shell particles. Dark blobs
  denote known momenta, light blobs unknown momenta.
  Arrows indicate the momentum flow, i.e.\ the order in which unknown
  momenta are determined from known ones. The $\bar{D}$-vertex
  corresponds to overall momentum conservation.\label{fig:ps_gen_blocks}}
}
\newcommand{\insertwgtvertices}{
\myfigure{tb}{
  \begin{picture}(140,121) (45,-45)
    \SetWidth{0.5}
    \SetColor{Black}
    \ArrowLine(165,61)(120,26)
    \ArrowLine(75,61)(120,26)
    \ArrowLine(120,26)(120,-29)
    \GOval(165,61)(13,13)(0){0.666}
    \GOval(75,61)(13,13)(0){0.666}
    \COval(120,26)(19,19)(45.0){Black}{White}
    \GOval(120,-29)(13,13)(0){0.888}
    \Text(75,61)[]{\Black{$\rho$}}
    \Text(165,61)[]{\Black{$\pi\setminus\rho$}}
    \Text(120,-29)[]{\Black{$\pi$}}
    \Text(120,26)[]{\Black{$\hat{S}_{\pi}^{\,\rho,\pi\setminus\rho}$}}
  \end{picture}  
  \begin{picture}(140,121) (45,-45)
    \SetWidth{0.5}
    \SetColor{Black}
    \DashArrowLine(165,61)(120,16){5}
    \ArrowLine(75,61)(120,16)
    \ArrowLine(120,16)(75,-29)
    \ArrowLine(165,-29)(120,16)
    \GOval(75,61)(13,13)(0){0.666}
    \GOval(165,61)(13,13)(0){0.666}
    \COval(120,16)(19,19)(45.0){Black}{White}
    \Text(75,61)[]{\Black{$\pi$}}
    \Text(165,61)[]{\Black{$\overline{\alpha b\pi}$}}
    \GOval(75,-29)(13,13)(0){0.888}
    \Text(165,-29)[lt]{\Black{$\hspace*{1ex}b$}}
    \Text(75,-29)[]{\Black{$\alpha$}}
    \Text(120,16)[]{\Black{$\hat{T}_{\alpha,b}^{\,\pi,\overline{\alpha b\pi}}$}}
  \end{picture}
  \begin{picture}(140,121) (45,-45)
    \SetWidth{0.5}
    \SetColor{Black}
    \ArrowLine(120,61)(120,10)
    \ArrowLine(120,6)(75,-29)
    \ArrowLine(165,-29)(120,6)
    \GOval(120,61)(13,13)(0){0.666}
    \COval(120,6)(19,19)(45.0){Black}{White}
    \Text(120,61)[]{\Black{$\overline{\alpha b}$}}
    \GOval(75,-29)(13,13)(0){0.888}
    \Text(165,-29)[lt]{\Black{$\hspace*{1ex}b$}}
    \Text(75,-29)[]{\Black{$\alpha$}}
    \Text(120,6)[]{\Black{$\hat{D}_{\alpha,b}$}}
  \end{picture}}{
  Basic decay vertices for weight calculation. Dark blobs
  denote potentially nontrivial known weights, light blobs
  weights to be determined. Arrows indicate the weight 
  flow, i.e.\ the order in which unknown weights are determined 
  from known ones. The $\hat{D}$-vertex corresponds to overall 
  momentum conservation.\label{fig:ps_weight_blocks}}
}
\newcommand{\insertexample}{
\begin{figure}[t]
\begin{center}
  \eeqqga$\rightarrow\quad$
  \begin{picture}(60,60)(0,25)
    \SetWidth{0.5}
    \SetColor{Black}
    \Line(30,30)(51,8)
    \Line(30,30)(5,5)
    \Line(5,55)(30,30)
    \DashLine(55,55)(30,30){2}
    \GOval(5,5)(8,8)(0){0.882}
    \COval(30,30)(12,12)(45.0){Black}{White}
    \GOval(5,55)(8,8)(0){0.882}
    \GOval(55,55)(8,8)(0){0.882}
    \Text(5,5)[]{\Black{$\sst a$}}
    \Text(55,5)[]{\Black{$\sst b$}}
    \Text(5,55)[]{\Black{$\sst 1$}}
    \Text(55,55)[]{\Black{$\sst 23$}}
    \Text(30,30)[]{\Black{$\sst T_{a,b}^{\,1,23}$}}
  \end{picture}$\quad\otimes\;$
  \begin{picture}(60,60)(0,25)
    \SetWidth{0.5}
    \SetColor{Black}
    \Line(30,24)(51,8)
    \Line(30,24)(5,5)
    \Line(30,55)(30,24)
    \GOval(5,5)(8,8)(0){0.882}
    \COval(30,24)(12,12)(45.0){Black}{White}
    \GOval(30,55)(8,8)(0){0.882}
    \Text(5,5)[]{\Black{$\sst a1$}}
    \Text(55,5)[]{\Black{$\sst b$}}
    \Text(30,55)[]{\Black{$\sst 23$}}
    \Text(30,24)[]{\Black{$\sst D_{a1,b}$}}
  \end{picture}$\otimes\quad P_{23}\;\otimes\;$
  \begin{picture}(60,60)(0,25)
    \SetWidth{0.5}
    \SetColor{Black}
    \Line(30,36)(55,55)
    \Line(30,36)(5,55)
    \Line(30,5)(30,36)
    \GOval(5,55)(8,8)(0){0.882}
    \GOval(55,55)(8,8)(0){0.882}
    \COval(30,36)(12,12)(45.0){Black}{White}
    \GOval(30,5)(8,8)(0){0.882}
    \Text(5,55)[]{\Black{$\sst 2$}}
    \Text(55,55)[]{\Black{$\sst 3$}}
    \Text(30,5)[]{\Black{$\sst 23$}}
    \Text(30,36)[]{\Black{$\sst S_{23}^{\,2,3}$}}
  \end{picture}\\
  \eeqqgb$\rightarrow\quad$
  \begin{picture}(60,60)(0,25)
    \SetWidth{0.5}
    \SetColor{Black}
    \Line(30,30)(51,8)
    \Line(30,30)(5,5)
    \Line(5,55)(30,30)
    \DashLine(55,55)(30,30){2}
    \GOval(5,5)(8,8)(0){0.882}
    \COval(30,30)(12,12)(45.0){Black}{White}
    \GOval(5,55)(8,8)(0){0.882}
    \GOval(55,55)(8,8)(0){0.882}
    \Text(5,5)[]{\Black{$\sst a$}}
    \Text(55,5)[]{\Black{$\sst b$}}
    \Text(5,55)[]{\Black{$\sst 23$}}
    \Text(55,55)[]{\Black{$\sst 1$}}
    \Text(30,30)[]{\Black{$\sst T_{a,b}^{\,23,1}$}}
  \end{picture}$\quad\otimes\;$
  \begin{picture}(60,60)(0,25)
    \SetWidth{0.5}
    \SetColor{Black}
    \Line(30,24)(51,8)
    \Line(30,24)(5,5)
    \Line(30,55)(30,24)
    \GOval(5,5)(8,8)(0){0.882}
    \COval(30,24)(12,12)(45.0){Black}{White}
    \GOval(30,55)(8,8)(0){0.882}
    \Text(5,5)[]{\Black{$\sst a23$}}
    \Text(55,5)[]{\Black{$\sst b$}}
    \Text(30,55)[]{\Black{$\sst 1$}}
    \Text(30,24)[]{\Black{$\sst D_{a23,b}$}}
  \end{picture}$\otimes\quad P_{23}\;\otimes\;$
  \begin{picture}(60,60)(0,25)
    \SetWidth{0.5}
    \SetColor{Black}
    \Line(30,36)(55,55)
    \Line(30,36)(5,55)
    \Line(30,5)(30,36)
    \GOval(5,55)(8,8)(0){0.882}
    \GOval(55,55)(8,8)(0){0.882}
    \COval(30,36)(12,12)(45.0){Black}{White}
    \GOval(30,5)(8,8)(0){0.882}
    \Text(5,55)[]{\Black{$\sst 2$}}
    \Text(55,55)[]{\Black{$\sst 3$}}
    \Text(30,5)[]{\Black{$\sst 23$}}
    \Text(30,36)[]{\Black{$\sst S_{23}^{\,2,3}$}}

    \DashLine(-40,-10)(-40,170){2}
    \DashLine(-40,170)(70,170){2}
    \DashLine(70,170)(70,-10){2}
    \DashLine(70,-10)(-40,-10){2}
  \end{picture}\\[2ex]
  \myfigcaption{0.95\textwidth}{
    Correspondence between Feynman diagrams and building blocks of
    the phase space for the process $q\bar{q}\to e^+e^-g$. The terms
    in the dashed box arise from both diagrams and have to be evaluated
    only once when computing the phase space weight.
    \label{fig:eeqqg_channels}}
  \end{center}
\end{figure}
}

In this section we present two new methods for integrating over the phase 
space. Both of them are designed to cope especially with large numbers of 
outgoing  particles. The first method is a fully general approach and 
makes use of the standard multi-channel technique~\cite{Kleiss:1994qy} 
in a recursive fashion, i.e.\ the phase space 
sampling fits the method of generating the corresponding matrix element. 
The second method is designed for QCD and QCD-associated processes and
employs the phase space generator 
\Haag~\cite{vanHameren:2002tc} in conjunction with a new prescription for 
coupling colour and momentum sampling and the multi-channel technique.

\mysubsection{Recursive algorithm for phase space integration}
\label{sec:recursive_phasespace}
One of the most efficient general approaches to sample the phase space 
of multi-particle processes is, to employ a multi-channel method according 
to Ref.~\cite{Kleiss:1994qy} with each of the single channels corresponding
to the pole structure of a certain Feynman diagram. However, for large 
numbers of diagrams this is clearly not the method of choice. 
In the following we will therefore focus on the recursive relations
for phase space generation proposed in Ref.~\cite{Byckling:1969sx}. We
construct a separate multi-channel for each possible subamplitude
on the flight according to the propagator structure and use 
\Vegas~\cite{Lepage:1980dq} 
to optimise the integration over propagator masses and polar
angles in decays. The obvious drawback of this procedure is evident:
It relies heavily on the assumption that the matrix element factorises
according to its propagator structure. However, it is a
generalisable way to tame the rather factorial growth in the number
of phase space channels encountered in conventional approaches~\cite{
  Kanaki:2000ey,*Papadopoulos:2000tt,*Cafarella:2007pc,
  Krauss:2001iv,Maltoni:2002qb,*Alwall:2007st}. 
If we take the prescription serious, we can factorise the full phase space
weight such that it can be computed in a recursive fashion
corresponding to how the matrix element is evaluated.
It turns out that this procedure gives an excellent performance, 
cf.\ Sec.~\ref{sec:results}.

\mysubsubsection{Brief review of phase space factorisation}
\label{sec:psi_intro}
In the following we consider a $2\to n$ scattering process and denote 
incoming particles by $a$ and $b$ and outgoing particles by $1\ldots n$.
The corresponding $n$-particle differential phase space element reads
\begin{equation}
  \done\Phi_n\rbr{a,b;1,\ldots,n}=
    \sbr{\,\prod\limits_{i=1}^n\frac{\done^4 p_i}{\rbr{2\pi}^3}\,
    \delta\rbr{p_i^2-m_i^2}\Theta\rbr{p_{i0}}\,}\,
    \rbr{2\pi}^4\delta^{(4)}\rbr{p_a+p_b-\sum_{i=1}^n p_i}\;,
\end{equation}
where $m_i$ are the on-shell masses of outgoing particles.
Following Ref.~\cite{James:1968gu}, the full phase space may be
factorised according to
\begin{equation}\label{eq:split_ps}
  \done\Phi_n\rbr{a,b;1,\ldots,n}=
    \done\Phi_{n-m+1}\rbr{a,b;\pi,m+1,\ldots,n}\,\frac{\done s_\pi}{2\pi}\,
    \done\Phi_m\rbr{\pi;1,\ldots,m}\;,
\end{equation}
where $\pi=\cbr{1,\ldots,m}$ corresponds to a set of particle indices, 
similar to Sec.~\ref{sec:sm}. Generally in this section, Greek letters 
denote a subset of all possible indices. Overlined letters denote the 
missing subset, i.e.\ $\overline{\alpha}=\{a,b,1,\ldots,n\}\setminus\alpha$ 
for all $\alpha\subset \{a,b,1,\ldots,n\}$. Similarly, $\overline{\pi_1\pi_2}
=(\,\{a,b,1,\ldots,n\}\setminus\pi_1)\setminus\pi_2$, etc. 
Equation~\eqref{eq:split_ps} allows to decompose the complete phase space
into building blocks corresponding to the $t$- and $s$-channel decay processes
    $T_{\alpha,b}^{\,\pi,\overline{\alpha b\pi}}=\;
      \done\Phi_{2}\rbr{\alpha,b;\pi,\overline{\alpha b\pi}}$ 
and    $S_{\pi}^{\,\rho,\pi\setminus\rho}=\;
      \done\Phi_{2}\rbr{\pi;\rho,\pi\setminus\rho}$
and the $s$-channel production process $D_{\alpha,b}$
, cf.\ Fig.~\ref{fig:ps_gen_blocks}.
We refer to these objects as phase space vertices, while the integral 
$P_\pi=\done s_\pi/2\pi$, introduced in Eq.~\eqref{eq:split_ps}, will be called 
a phase space propagator. We use the same notation as for the propagators
in Sec.~\ref{sec:sm} to highlight the close correspondence between matrix element
computation and phase space generation. In the algorithm presented here, 
only timelike propagators are employed.
\insertgenvertices

The phase space vertices are used differently in the case of weight calculation 
and phase space generation. Consider the $t$-channel decay. 
If a phase space point is to be generated, the new final state momenta $p_\pi$ and 
$p_{\overline{\alpha b\pi}}$ are determined from the known initial state
momenta $p_\alpha$ and $p_b$. If a weight needs to be computed, the new weight 
$w_\alpha^{(b)}$ is determined from the vertex weight and the input weights
$w_\pi$ and $w_{\overline{\alpha b\pi}}$. The corresponding situations
are depicted in Figs.~\ref{fig:ps_gen_blocks} and~\ref{fig:ps_weight_blocks}, 
respectively. The basic building blocks of phase space integration 
are summarised as follows
\begin{equation}\label{eq:ps_building_blocks}
  \begin{split}
    P_{\pi}&=\left\{\begin{array}{cc}1&
      \text{if $\pi$ or $\bar{\pi}$ external}\\
      \displaystyle\frac{\done s_\pi}{2\pi}&
      \text{else}\end{array}\right.\;,\\
    S_{\pi}^{\,\rho,\pi\setminus\rho}&=
     \frac{\lambda\rbr{s_\pi,s_\rho,s_{\pi\setminus\rho}}}{
     16\pi^2\,2\,s_{\pi}}\;\done\cos\theta_\rho\,\done\phi_\rho\;,\\
    T_{\alpha,b}^{\,\pi,\overline{\alpha b\pi}}&=
     \frac{\lambda\rbr{s_{\alpha b},s_\pi,
       s_{\;\overline{\alpha b\pi}}}}{16\pi^2\,2s_{\alpha b}}\;
     \done\cos\theta_\pi\,\done\phi_\pi\;,\\
    D_{\alpha,b}&=\rbr{2\pi}^4\,\done^4 p_{\,\overline{\alpha b}}\;
      \delta^{(4)}\rbr{p_\alpha+p_b
      -p_{\,\overline{\alpha b}}}\;.
  \end{split}
\end{equation}
Here we have introduced the triangular function
\begin{equation}
  \lambda\rbr{s_a,s_b,s_c}=\sqrt{\rbr{s_a-s_b-s_c}^2-4s_bs_c}
\end{equation}
Note that even since $\alpha$ might correspond to an off-shell internal 
particle, $b$ always indicates a fixed external incoming particle. This is
essential in all further considerations and allows reusing weight factors
in the Monte Carlo integration, just as currents are reused in the matrix
element computation. The functions corresponding to 
$S_{\pi}^{\,\rho,\pi\setminus\rho}$ and 
$T_{\alpha}^{\,\pi,\overline{\alpha b\pi}}$ 
are in fact identical, since they represent a solid angle integration. 
In practice however we choose the different sampling strategies proposed  
in Ref.~\cite{Byckling:1969sx}. The $s$-channel production vertex $D_{\alpha,b}$
is only needed for bookkeeping, since it corresponds to overall momentum 
conservation and the associated overall weight factor $(2\pi)^4$.

\mysubsubsection{A simple example}
We illustrate in this section how a recursive phase space generator for 
the process $q\bar{q}\to e^+e^-g$ can be constructed, based on the 
diagrammatic structure of the integrand. Figure~\ref{fig:eeqqg_channels}
depicts the translation of the corresponding Feynman diagrams into related
building blocks of the phase space. The standard procedure to define an
integrator consists of constructing one integration channel per line of
Fig.~\ref{fig:eeqqg_channels} and joining these channels in a multi-channel.
Because it is based on full diagrams, this strategy cannot be implemented 
in a recursive fashion and we have to modify it. Consider first,
which tasks have to be performed for each phase space point.
\insertwgtvertices

To generate momenta, one starts with the $s$-channel propagator $P_{23}$.
Then, depending on what yields the better performance, the $t$-channel decay 
$T_{a,b}^{\,1,23}$ or $T_{a,b}^{\,23,1}$ and finally the $s$-channel decay 
$S_{23}^{\,2,3}$ are constructed, leading to the final state momenta
$p_1\ldots p_3$. Note again that $D$-type vertices are just for bookkeeping
at this point. They simply imply overall momentum conservation.
When computing the phase space weight, the order of treating vertices
can be reversed because all corresponding momenta are known. Therefore the 
weights for the decay $S_{23}^{\,2,3}$ and the propagator $P_{23}$ 
are computed first, followed by the weights for $T_{a,b}^{\,1,23}$ and 
$T_{a,b}^{\,23,1}$. It is obvious that the weights $\hat{P}_{23}$ and 
$\hat{S}_{23}^{\,2,3}$ are unique and therefore have to be computed only
once, although arising in both lines of Fig.~\ref{fig:eeqqg_channels}.
We refer to this feature as the ``weight flow'', which is directed 
from the final state particles and the right beam, particle $b$, 
towards the left beam, particle $a$. This generalises to arbitrary processes, 
provided that the right beam particle is kept fix, i.e.\ $t$-channel indices 
always combine only $a$ and external indices, as indicated in 
Figs.~\ref{fig:ps_gen_blocks} and~\ref{fig:ps_weight_blocks}. 
It allows to compute the full phase space weight recursively, in a manner
similar to Eq.~\eqref{eq:sm_general_recursion}, which implies in particular, 
that at most the same growth is induced in the matrix element and the 
phase space weight computation.

Let us illustrate this new procedure using the above example. 
The difference with respect to the standard approach is how multi-channels
are defined. Following the weight flow, in the first step of
the recursion we construct a multi-channel for the phase space element 
$\done\Phi_2\rbr{\{23\};2,3}$. Of course, since particles $2$ and $3$ 
are external, this multi-channel consists of one single channel only, 
which is the $s$-channel decay $S_{23}^{\,2,3}$.
It has therefore no additional parameters. In the second step we construct
a multi-channel for the full phase space $\done\Phi_3\rbr{a,b;1,2,3}$, which
receives contributions from the two $t$-channels $T_{a,b}^{1,23}$ and 
$T_{a,b}^{\,23,1}$. Each of them can be assigned a multi-channel weight $w$,
which eventually yields the overall weight 
\begin{equation}\label{eq:example_mc_dy1j}
  (2\pi)^4\,F^{-1}\sbr{\,\frac{
    w_{a,b}^{1,23}F\sbr{\hat{T}_{a,b}^{\,1,23}\hat{P}_{23}\hat{S}_{23}^{\,2,3}}
    +w_{a,b}^{23,1}F\sbr{\hat{T}_{a,b}^{\,23,1}\hat{P}_{23}\hat{S}_{23}^{\,2,3}}}{
    w_{a,b}^{1,23}+w_{a,b}^{23,1}}\,}\;,
\end{equation}
Here $\hat{P}$ denotes the propagator weight, and $F$ is a generalised 
mean function, see next section. The overall factor $(2\pi)^4$ arises 
from the $D$-type vertices. 

We are left with the task to determine the necessary building blocks 
of the phase space using information from the matrix element. This 
turns out to be extremely simple since we now in fact have a phase space
weight recursion of the form of Eq.~\eqref{eq:sm_general_recursion} 
(cf.\ Eqs.~\eqref{eq:ps_recursion_s} and~\eqref{eq:ps_recursion_t}),
where multi-channels are associated with intermediate $s/t$-channel
propagators. Therefore each intermediate current in the matrix element 
implies a separate multi-channel in the corresponding integrator 
and each vertex implies a decay vertex associated with a single channel.
Because of this correspondence the default sampling strategy for 
$s$-channel propagator masses can be chosen according to the type of 
intermediate particle, i.e.\ a Breit--Wigner-like distribution 
for massive, unstable particles and a $1/s^\alpha$-type distribution 
for massless particles. These distributions as well as the polar angle
distributions in decays (cf.\ Eq.~\eqref{eq:ps_building_blocks}) 
are further refined during the integration using \Vegas.

%The above example shows how a process specific recursive phase space 
%generator can be constructed in an automated way. Following the weight flow,
%multi-channels are built up for every different intermediate state
%and the full initial state using matrix element information to determine
%the respective building blocks.
\insertexample

\mysubsubsection{Formulation of the recursive algorithm}
In this subsection we derive the general algorithm for the recursive 
phase space integrator. We employ the notation of Sec.~\ref{sec:psi_intro}.
Recursive relations for phase space integration in terms 
of the quantities introduced ibidem can be defined through
\begin{equation}\label{eq:ps_recursion_one}
  \begin{split}
    \done\Phi_S\rbr{\pi}&=\left.S_{\pi}^{\pi_1,\pi_2}\;
      P_{\pi_1}\,\done\Phi_S\rbr{\pi_1}\,
      P_{\pi_2}\,\done\Phi_S\rbr{\pi_2}\;
      \vphantom{\done\Phi_T^{(b)}}\right|_{\;
        \rbr{\pi_1,\pi_2}\in\mc{OP}\rbr{\pi}}\;,\\
    \done\Phi_T^{(b)}\rbr{\alpha}&=\left.T_{\alpha,b}^{\pi_1,\pi_2}\;
      P_{\pi_1}\,\done\Phi_S\rbr{\pi_1}\,
      P_{\pi_2}\,\done\Phi_T^{(b)}\rbr{\alpha\pi_1}\;\right|_{\;
        \rbr{\pi_1,\pi_2}\in\mc{OP}\rbr{\overline{\alpha b}}}
      +D_{\alpha,b}\;\done\Phi_S\rbr{\overline{\alpha b}}\;.
  \end{split}
\end{equation}
The above equations correspond to selecting one possible splitting
of the multi-index $\pi$ or $\overline{\alpha b}$ per phase space point. 
We can improve the 
integration procedure by forming an average over all possible splittings
in the spirit of a multi-channel. Let $F$ be a generalised mean function. 
We can then use the $F$-mean to define
\begin{equation}\label{eq:ps_recursion_s}
  \begin{split}
    \done\Phi_S\rbr{\pi}&=F^{-1}\left[\;
      \rbr{\sum\limits_{\rbr{\pi_1,\pi_2}\in\mc{OP}\rbr{\pi}}\,
      \omega_\pi^{\pi_1,\pi_2}\;}^{-1}\right.\\
      &\qquad\left.\times
      \sum\limits_{\rbr{\pi_1,\pi_2}\in\mc{OP}\rbr{\pi}}\,
      \omega_\pi^{\pi_1,\pi_2}\,F\sbr{\;S_{\pi}^{\pi_1,\pi_2}\;
        P_{\pi_1}\,\done\Phi_S\rbr{\pi_1}\,
        P_{\pi_2}\,\done\Phi_S\rbr{\pi_2}
        \vphantom{\done\Phi_T^{(b)}}\;}\;\right]\;,
  \end{split}
\end{equation}
\begin{equation}\label{eq:ps_recursion_t}
  \begin{split}
    \done\Phi_T^{(b)}\rbr{\alpha}&=F^{-1}\left[\;
      \rbr{\;\omega_{\alpha,b}\;+\sum\limits_{\rbr{\pi_1,\pi_2}\in\mc{OP}
        \rbr{\overline{\alpha b}}}\,
      \omega_\alpha^{\pi_1,\alpha\pi_1}\;}^{-1}\right.\\
      &\qquad\qquad\left.\left(\;
        \omega_{\alpha,b}\;F\sbr{\vphantom{\Phi_T^{(b)}}\;
        D_{\alpha,b}\;\done\Phi_S(\overline{\alpha b})\;}\;+
      \sum\limits_{\rbr{\pi_1,\pi_2}\in\mc{OP}
        \rbr{\overline{\alpha b}}}\,
      \omega_\alpha^{\pi_1,\alpha\pi_1}\,\right.\right.\\
      &\qquad\qquad\qquad\qquad\left.\left.
      \vphantom{\sum_{\mc{OP}\rbr{\overline{\alpha b}}}}\times F\sbr{\;
        T_{\alpha,b}^{\pi_1,\pi_2}\;P_{\pi_1}\,\done\Phi_S\rbr{\pi_1}\,
        P_{\pi_2}\,\done\Phi_T^{(b)}\rbr{\alpha\pi_1}\;}\right)\;\right]\;.
  \end{split}
\end{equation}
In this context we define the one- and no-particle phase space
\begin{equation}
  \begin{split}
    \done\Phi\rbr{i}&=1\;,\\
    \done\Phi\rbr{\emptyset}&=0\;.
  \end{split}
\end{equation} 
The function $\omega$ corresponds to a vertex-specific weight which may be 
adapted to optimise the integration procedure, see Ref.~\cite{Kleiss:1994qy}. 
The second sums run over all possible $S$- and $T$-type vertices which have 
a correspondence in the matrix element. The full differential phase space 
element is given by
\begin{equation}
  \done\Phi_n\rbr{a,b;1,\ldots,n}=\done\Phi_T\rbr{a}\;.
\end{equation}
Note that Eqs.~\eqref{eq:ps_recursion_s} and~\eqref{eq:ps_recursion_t} 
in the form stated above are {\it not} suited to generate the sequence 
of final state momenta.  To do so one rather has to employ the following 
algorithm, which corresponds to a reversion of the recursion and respects 
the weight factors $w$ introduced above.
\begin{itemize}
\item From the set of possible vertices connecting currents in the 
  matrix element, choose a sequence connecting all external particles
  in the following way:
  \begin{enumerate}
  \item Start with the set of indices $\pi=\cbr{b,1,\ldots,n}$,\\
    corresponding to the unique external current of index $a$.
  \item\label{ps_gen_rec_step_2} From the set of possible phase space 
    vertices connecting to $\pi$ select one according to an on the flight
    constructed multi-channel employing the weights $w$.\footnote{
      Note that in this context weights have to be normalised
      to unity on the flight.} 
    If $\pi$ is a single index, stop the recursion.
  \item According to the selected vertex, split $\pi$ into the subsets
    $\pi_1$ and $\pi_2$. Repeat step~\ref{ps_gen_rec_step_2} for these 
    subsets.
  \end{enumerate}
\item Fore each vertex, make use of the fact that $\pi$ is equivalent 
  to $\overline{\pi}$ and adjust the indices in an appropriate way 
  for momentum generation.
  That is if any $\pi$ contains $b$ and other indices, replace $\pi$ by 
  $\overline{\pi}$.
\item Order $\bar{T}$-type vertices ascending and $\bar{S}$-type vertices 
descending in the number of external indices connected to initial states.
\item Generate the corresponding momenta starting with 
  $\bar{T}$-type vertices.
\end{itemize}
Even though $T$-type vertices depend on $b$, since $b$ is fixed throughout 
the computation of one phase space point we obtain no expressions 
depending on more than two particle indices. This induces the same growth 
of computational complexity in both the hard matrix elements and the phase 
space and makes the above algorithm well suited for integration of processes 
with large final state multiplicity. In the following we refer to it
as the Recursive Phase space Generator (RPG).

\mysubsubsection{Implementation details}
Since the phase space weight computation, Eq.~\eqref{eq:ps_recursion_s}
obeys a recursion similar to those of the matrix element calculation, 
Eq.~\eqref{eq:sm_general_recursion}, it is straightforward to implement
this weight computation into a numerical program along the lines
of Sec.~\ref{sec:implementation_me}. The same techniques described
for the multi-threading of matrix element calculations can be implemented
for the phase space weight. In the multi-threaded version
of \Comix, this weight is computed in parallel to the matrix element, 
which further reduces the net computation time 
if enough resources are available.

\mysubsection{Colour sampling}
\label{sec:colour_sampling}
For QCD and QCD associated processes with a large number of external legs, 
it becomes unfeasible to compute colour-summed scattering amplitudes.
Instead the better strategy is to sample over external colour assignments 
in a given representation of SU(3). According to 
Eqs.~\eqref{eq:color_decomposition_fundamental} - 
\eqref{eq:color_decomposition_colorflow}, this selects a set of colour-ordered 
amplitudes which contribute to the corresponding point in colour space.
This set is typically strongly reduced compared to the full set of
partial amplitudes. The issue has been studied in Ref.~\cite{Maltoni:2002mq} 
for the fundamental representation decomposition, the adjoint representation 
decomposition and the colour flow decomposition, which has been presented therein. 
The conclusion is that the colour flow decomposition is the method best suited 
for sampling over colour assignments if the number of external partons is large, 
i.e.\ it provides the slowest growth in the average number of partial amplitudes 
per non-vanishing colour assignment. Also it has been exemplified for recursive 
calculations in Ref.~\cite{Duhr:2006iq}, that the colour flow decomposition 
is advantageous, since no computational intensive matrix multiplications 
have to be performed. We therefore employ this prescription throughout \Comix.

In the following we focus on an $n$-gluon scattering process. However,
the presented ideas and algorithms are straightforward to generalise for
arbitrary sets of colour octet objects, such as e.g.\ quark-antiquark pairs.
In the colour flow decomposition each external gluon is labeled by a colour
index $i$ and an anti-colour index $\bar\jmath$. The colour assignment for an 
$n$-gluon scattering is thus given by selecting each index $i_1,\ldots i_n$ 
and $\bar\jmath_1,\ldots \bar\jmath_n$ out of three values $\rbr{R,G,B}$ and
$\rbr{\bar R,\bar G,\bar B}$.

\mysubsubsection{Determination of colour flows from colour assignments}
\label{sec:ca2cf}
A specific colour flow, and thus an ordering in the sense of a 
colour-ordered amplitude, is specified by a permutation 
\begin{equation}\label{eq:cfperm}
  \vec{\sigma}=\rbr{1,\sigma_2,\sigma_3,\ldots,\sigma_n}\in S_{n-1}\;
\end{equation}
of external gluon indices. 
This colour flow contributes to a colour assignment, if
\begin{equation}\label{valid_cf_cond}
  \delta^{i_1 \bar\jmath_{\sigma_2}}
  \delta^{i_{\sigma_2} \bar\jmath_{\sigma_3}}
  \cdots\delta^{i_{\sigma_n} \bar\jmath_1}=1\;.
\end{equation}
It is thus easy to construct an algorithm which determines all valid 
colour flows from a given colour assignment.
\begin{enumerate}
  \item Set the first gluon index to $\sigma_1=1$. Let $k=2$.
  \item\label{colorflow_algo_step}
    Select one of the remaining gluon indices to be $\sigma_k$,
    such that $i_{\sigma_{k-1}}=\bar\jmath_{\sigma_k}$.
    If this is possible, let $k\to k+1$. Otherwise
    let $k\to k-1$, then repeat this step selecting a different $\sigma_k$.  
  \item
    If $k=n+1$ and $i_{\sigma_{n}}=\bar\jmath_{\sigma_1}$,
    a valid flow has been found.\\
    Otherwise continue with step~\ref{colorflow_algo_step}.
\end{enumerate}
By systematically selecting through all possible $\sigma_k$ in 
step~\ref{colorflow_algo_step} all valid colour flows are determined.

\mysubsubsection{Selection of colour assignments}
\label{sec:selcolour}
The simplest way of choosing a colour assignment is accomplished
by randomly selecting the $2n$ colours for the $i$- and $\bar\jmath$-indices. 
Each colour is chosen with an equal probability, leading to a weight of $3^{2n}$.
However, only a small fraction of those assignments will have at least one valid
colour flow. A trivial (but not sufficient) condition for non-vanishing
amplitudes is, that the number of $i$-indices carrying the colour $R$ ($G$,$B$)
must be equal to the number of $\bar\jmath$-indices carrying the corresponding 
anticolour.

We thus propose a more efficient way to determine colour configurations.
\begin{enumerate}
\item The $n$ $i$-indices are selected randomly in $(R,G,B)$.
\item A permutation $\vec\sigma=\rbr{\sigma_1,\ldots,\sigma_n}$ 
  of $n$ particles is selected randomly with a uniform weight. 
  The anticolours of the $\bar\jmath$-indices are then given by
  $\bar\jmath_k=\overline{i_{\sigma_k}}$, for $k=1,\ldots,n$.
\item Each colour assignment is weighted by
  \begin{equation}\label{eq:cioweight}
    w\,=\;3^n\,\frac{n!}{n_R!\,n_G!\,n_B!}\;,
  \end{equation}
  where $n_R$, $n_G$ and $n_B$ are the multiplicities of $i$-indices\\ 
  carrying the colours $R$, $G$ and $B$, respectively. 
\end{enumerate}
Clearly, assignments generated by this algorithm will always fulfil the trivial 
condition mentioned above. Moreover, the weight is roughly proportional to the number
of possible colour flows and thus already corresponds to some extent
to the expected cross section for this colour configuration.

\mysubsubsection{A simple example}
To illustrate the colour sampling in the colour flow decomposition we consider
a five gluon scattering process. The random selection of a colour configuration
using the improved algorithm may return the following $i$-indices:
\begin{equation}
i_1=R\,,\;\;\;i_2=R\,,\;\;\;i_3=G\,,\;\;\;i_4=G\,,\;\;\;i_5=B\,.
\end{equation}
The $\bar\jmath$-indices are fixed by a randomly chosen permutation, say
$\vec{\sigma}=\rbr{4,1,2,5,3}$:
\begin{equation}
\bar\jmath_1=\bar G\,,\;\;\;\bar\jmath_2=\bar R\,,\;\;\;\bar\jmath_3=\bar R\,,\;\;\;
\bar\jmath_4=\bar B\,,\;\;\;\bar\jmath_1=\bar G\,.
\end{equation}
For this assignment the only colour flow that satisfies Eq.~\eqref{valid_cf_cond}
is given by the permutation $\vec{\sigma}=\rbr{1,4,5,3,2}$.

\mysubsection{Combined colour-momentum integration techniques}
\label{sec:col_mom_integrator}
Generally the peaking behaviour of the colour-sampled differential cross section 
is rather complex within the phase space and strongly different for different 
colour assignments.
The idea must thus be to construct integrators specific for a given 
colour assignment, based on the knowledge of contributing partial amplitudes.
One can for example think of a variant of the algorithm described in 
Sec.~\ref{sec:recursive_phasespace}, where the basic building blocks of the phase space 
are either available or not, depending whether there is a corresponding non-vanishing 
coloured current present in the matrix element. However, in practice this choice does not 
lead to any significant improvement of the integration behaviour of the RPG and we thus 
refrain from promoting this method.\footnote{Note that this is not a statement about the
  integration behaviour of the RPG itself, but only about a possible coupling of colour
  and momentum sampling using the RPG.} Instead we present a second type of integrator, 
dedicated to be used with QCD and QCD associated processes, which is based on the
\Haag algorithm~\cite{vanHameren:2002tc}.
As before we concentrate on purely gluonic processes.

\mysubsubsection{Integration of partial amplitudes and colour configurations}
\label{sec:csi}
As a basic building block we use the \Haag-integrator, which generates momenta 
distributed proportional to a QCD antenna function~\cite{vanHameren:2002tc},
\begin{equation}
\label{eq:antenna}
  A_{n}(p_1,p_2,...,p_{n})=\frac{1}{
    (p_1 p_2)(p_2 p_3)...(p_{n-1} p_{n})(p_{n} p_1)}.
\end{equation}
Details on our implementation of the algorithm and improvements to the original version 
are given in Ref.~\cite{Gleisberg:2008ft}.
Single \Haag-channels provide efficient integrators for squared partial amplitudes
associated with a given colour flow, 
both labeled by the same permutation $\vec\sigma$, Eq.~(\ref{eq:cfperm}). 
For the \Haag-channel the permutation corresponds to the order of momenta
in the antenna function.
As for the RPG we again obtain a close correspondence between the matrix element
and the phase space generation, now at the level of partial amplitudes.

The cross section for a single colour assignment is given by the squared sum
of partial amplitudes associated with all valid colour flows. Ignoring
the interferences between the amplitudes in the context of the phase space setup, a dedicated
integrator can be constructed by combing the corresponding \Haag-channels for
each flow in a multi-channel integrator.
With growing number of external particles, however, one faces the following problem:
Although the average number of contributing colour flows per 
colour assignment is relatively low in the colour flow  decomposition, the maximal number 
grows factorially. Thus it quickly becomes impossible to store all data 
associated with the multi-channels defined above, i.e.\ the contributing \Haag-channels 
and the internal weights. The situation gets even worse if the sampling over all 
colour assignments is taken into account, because the number of possible assignments 
grows exponentially with the number of external particles. 
The solution to this is not to store any multi-channel parameters, 
but to generate the complete multi-channel on the flight.

A fast algorithm, as presented in Sec.~\ref{sec:ca2cf} to provide 
all colour flows from a colour assignment is essential for this approach: 
for a single phase space point one has to loop
three times over the list of all colour flows (which cannot be stored as well due to
the factorially growing maximal number of flows).
\begin{enumerate}
  \item To determine the normalisation of weights 
    within the multi-channel integrator. 
  \item To select a channel for generating a phase space point
    with a probability given by the relative weight $\alpha_k$, and
  \item To compute the multi-channel weight corresponding to this phase space point.
\end{enumerate}

\mysubsubsection{Optimisation techniques}
The proposed integrator contains a number of parameters which can be adjusted or 
adapted to reduce the variance during integration.
A multi-channel integrator dedicated to a specific colour assignment has the following 
degrees of freedom for optimisation:
\begin{itemize}
  \item \Vegas grids to refine individual \Haag-channels,
  \item Relative weights $\alpha_k$ in the multi-channel generator,
\end{itemize}
The sheer multiplicity of different channels and on-the-flight construction
of the integrator forbids an individual adaptation of all parameters.
However, their number can be greatly reduced by making use of the
symmetry among different \Haag-channels w.r.t. to permutations of the final state. 
All channels with the same relative positions of the initial state momenta within
the antenna can be determined from each other by a permutation of final state momenta.
This prevents the number of structurally different phase space channels from growing 
factorially with the number of particles and induces a linear growth only. 
Taking into account that the same symmetry holds for the partial amplitudes 
justifies to reuse the optimisation parameters among all channels of one kind. 
For later reference we label different types of \Haag-channels 
(and respective partial amplitudes) by the number of
final state momenta between the first and the second incoming momentum within
a certain antenna.

We achieve the best integration efficiency by performing the optimisation of the
free parameters prior to the actual integration:
The \Vegas grids of the \Haag-channels are adapted individually by
integrating corresponding single squared partial amplitudes over the allowed phase space. 
Using the above mentioned symmetry this has to be done only for one 
channel of each kind\footnote{ During this step the full result can not
    be determined since potential interferences between partial amplitudes 
    are ignored. However, it is sufficient for computing the leading $1/N_C$ 
    limit for $n$ gluon processes, using the fact that in the colour flow 
    decomposition (as well as in the fundamental representation decomposition)
    interferences are always subleading.}.
This technique not only speeds up the optimisation, it also provides a much cleaner 
environment for the adaptation of the \Vegas grids. At this stage a summation
over helicities is performed. Cross sections $\sigma_t$, given by the integration 
result from the channel of type $t$, are stored.

In the actual integration run
no further optimisation is performed. The channels are used as they emerged
from the above procedure, including the \Vegas-grid and a parameter $\alpha_k$, 
proportional to the cross section, $\sigma_t$, of the corresponding squared 
partial amplitude.

Best performance is achieved, if the colour assignment is selected 
with a probability proportional to the sum of cross sections of 
contributing squared partial amplitudes (as determined during 
the optimisation step), instead of the weight given by 
Eq.~\eqref{eq:cioweight}. To do so, the total normalisation for the 
new weight must be determined summing over all colour assignments.
For $n$-gluon processes this number is given by the following simple formula:
\begin{equation}
  N =\, (n-2)\,!\; 3^n\, \sum_{i=0}^{n-2} \sigma_{\min(i,n-i-2)}\;,
\end{equation}
where the $\sigma_{\min(i,n-i-2)}$ is the cross section of a squared partial 
amplitude of the type ``$\min(i,n-i-2)$''. The reweighting can be done 
by a simple hit-or-miss method.

For the integration run it is a matter of choice whether to sum or sample 
over helicities. All practical tests for up to the 11-gluon process favoured 
summation. Beyond that, however, it seems to become too costly to compute 
summed matrix elements, thus a sampling should be considered.

In the context of this work, we refer to the above algorithm as the 
Colour Sampling Integrator (CSI).

%% file: text/results.tex
\newcolumntype{a}[2]{>{\raggedleft}p{#1}@{}p{#2}}

In this section we present selected results generated with \Comix.
We focus on the special feature of this new generator, to be suitable 
in particular for computation of large multiplicity matrix elements.
A detailed comparison of integration times, compared to a dedicated
code using CSW vertex rules and the generator \Amegic can be found
in Ref.~\cite{Gleisberg:2008ft}.

\mysubsection{Helicity summation vs. helicity sampling}

Firstly we illustrate the effect of suitable matrix element
generation in the helicity summed mode of \Comix, see 
Sec.~\ref{sec:implementation_me}. Computation times for helicity summed 
and helicity sampled matrix elements in pure gluonic processes 
are compared in Tab.~\ref{tab:helicity_sum_sample}. The naive ratio between
the two is the number of possible helicity assignments of the respective
amplitude, $2^n-2(n+1)$, with $n$ the number of external gluons.
This naive ratio corresponds to computing the amplitude afresh for each 
of the different helicity assignments. Employing the ideas presented
in Sec.~\ref{sec:implementation_me}, however we find that this value
overestimates the real computational cost by up to a factor
of $\approx 7$. Obviously this statement is process dependent. The general
feature, however is that there is a gain when computing helicity
summed matrix elements. For the computation of cross sections 
this type of calculation might be preferred over the helicity sampled mode,
especially when using the phase space integration methods of the previous 
chapter, which are not designed for helicity sampling.

\mysubsection{Performance of the CSI and $2\to n$ gluon benchmarks}
In this subsection we present a comparison of gluon production cross sections
to illustrate both the performance of the CSI and the efficiency of the 
matrix element generation. We start with a fixed centre-of-mass energy.
The parameters are those of Refs.~\cite{Caravaglios:1998yr,Maltoni:2002mq}, 
i.e.\ $\alpha_S=0.12$ and
\begin{align}\label{eq:fabio_xsc}
  p_{Ti}&>60\;{\rm GeV}\;,&|\eta_i|&<2\;,&\Delta R_{ij}&>0.7\;,
\end{align}
for all final state gluons $i$ and pairs of gluons $i,j$. Integration results are
summarised in Tab.~\ref{tab:fabio_xsc}. We find perfect agreement with the 
results in the literature and give new predictions for the processes 
$gg\to 11g$ and $gg\to 12g$. Results have been generated with the CSI,
except for the $2\to 11$ and $2\to 12$ process, where RAMBO~\cite{Kleiss:1985gy} 
has been employed.
In order to examine the performance of the new phase space generator 
in a more realistic scenario, we investigate the same partonic processes
at the LHC and employ the Tevatron Run~II ${\rm k}_T$ 
algorithm~\cite{Blazey:2000qt}\footnote{ Note that we replace 
  $\Delta R_{ij}^2\to 2\,\sbr{\,\cosh\Delta\eta_{ij}-\cos\Delta\phi_{ij}\,}$ in order
  to match the Durham measure for final state clusterings.}
to define a cut on the multi-particle phase space. The respective results are
summarised in Tab.~\ref{tab:gluon_xsc}. We find that the CSI performs 
very well in both cases, even for large multiplicities, such that the 
respective cross sections can be computed with good precision.

Figures~\ref{fig:csi_conv_1} and~\ref{fig:csi_conv_2} show the convergence 
behaviour of the CSI for various gluon multiplicities. Since the computation of 
$2\to 8$ and $2\to 9$ gluon processes is quite cumbersome, it is worthwhile 
to switch to the helicity sampled mode in that case. Correspondingly we compare 
the performance of the CSI in helicity summed and helicity sampled mode in 
Fig.~\ref{fig:csi_conv_2}.

\mysubsection{Performance of the RPG and comparison with other generators}
We finally compare the performance of \Comix with those of other programs. 
All results presented in this section were obtained with the RPG
described in Sec.~\ref{sec:recursive_phasespace}. As references we use 
\Amegic~\cite{Krauss:2001iv} and \Alpgen~\cite{Mangano:2002ea}. The original setup 
for the comparison was established during the MC4LHC workshop~\cite{MC4LHC:2003aa}.
For a comprehensive listing of results from all participating projects,
see ibidem. Input parameters are given in Tab.~\ref{tab:mc4lhc_params}.
Cross sections are summarised in Tabs.~\ref{tab:mc4lhc_qcd}~-~\ref{tab:mc4lhc_vb} 
and~\ref{tab:mc4lhc_sub}. 

As pointed out in Sec.~\ref{sec:recursive_phasespace}, a drawback of the RPG 
is that it might not be able to adapt to certain peaks of the matrix element which 
correspond to specific diagrams. No significant disadvantage compared to
other generators can however be observed. A measure for the efficiency of a 
phase space generator is given by the ratio of the average over the maximal
weight $\abr{w}/w_{\rm max}$, i.e. the efficiency for generating events of
unit weight using a hit-or-miss method. As discussed in Ref.~\cite{Jadach:1999sf},
the maximum weight and thus this ratio is a numerically rather unstable quantity, 
often determined by very rare events in the high tail of the weight distribution.
In Tab.~\ref{tab:mc4lhc_effs} we therefore list the more stable quantity
$\abr{w}/w^\varepsilon_{\rm max}$, where the reduced maximum weight $w^\varepsilon_{\rm max}$ is 
defined such that $1-\abr{\min(w,w^\varepsilon_{\rm max})}/\abr{w}=\varepsilon\ll 1$.
It turns out that we achieve a reasonably good efficiency using the RPG, 
even for very large multiplicities. It can therefore be concluded that this
phase space generator is an excellent approach to tame the factorial growth 
of phase space channels while still maintaining an a priori adaptation to the
assumed peak structure of the integrand.

\mytable{p}{
  \begin{minipage}{0.85\textwidth}\begin{center}
  \begin{tabular}{|p{2cm}|a{1.2cm}{1.0cm}|a{1.1cm}{1.1cm}|
                     a{1.0cm}{0.6cm}|a{0.8cm}{0.8cm}|}\hline 
    Process\vphantom{\Large P} & \multicolumn{4}{c|}{Time [ ms / pt ] } & 
    \multicolumn{2}{c|}{} &\multicolumn{2}{c|}{} \\\cline{2-5}
    & \multicolumn{2}{c|}{sum} & \multicolumn{2}{c|}{sample} & 
    \multicolumn{2}{c|}{Ratio}&\multicolumn{2}{c|}{Gain}\\\hline
    $gg \to 2g$ & 0&.073  & 0&.025  & 2&.9 & 2&.1 \\
    $gg \to 3g$ & 0&.339  & 0&.060  & 5&.7 & 3&.5 \\
    $gg \to 4g$ & 1&.67   & 0&.149  & 11&  & 4&.5 \\
    $gg \to 5g$ & 8&.98   & 0&.427  & 21&  & 5&.3 \\
    $gg \to 6g$ & 49&.6   & 1&.39   & 36&  & 6&.6 \\
    $gg \to 7g$ & 298&    & 4&.32   & 69&  & 7&.1 \\
    $gg \to 8g$ & 1990&   & 13&.6   & 146& & 6&.9 \\
    $gg \to 9g$ & 13100&  & 43&.7   & 300& & 6&.7 \\
    $gg \to 10g$ & 96000& & 138&    & 695& & 5&.9 \\\hline 
  \end{tabular}\vspace*{5mm}\end{center}\end{minipage}}{
  Computation time for multi-gluon scattering matrix elements sampled over colour
  configurations. Displayed times are averages for a single evaluation
  of the colour dressed BG recursion relation, 
  when summing and sampling over helicity configurations, respectively. 
  Additionally in the last column, labeled `Gain' we give the inverse
  ratio of evaluation times multiplied by the naive ratio $2^n-2(n+1)$, where $n$
  is the number of external gluons. Numbers were generated on a 2.80~GHz 
  Pentium$^\text{\textregistered}$ 4 CPU.\label{tab:helicity_sum_sample}}

\mytable{p}{
  \begin{minipage}{0.85\textwidth}\begin{center}
  \begin{tabular}{|p{2cm}
    |a{0.3cm}{1.3cm}|a{0.3cm}{1.3cm}|a{0.3cm}{1.3cm}
    |a{0.3cm}{1.3cm}|a{0.3cm}{1.3cm}|}\hline
    gg $\to$ ng\vphantom{\Large P} & \multicolumn{10}{c|}{Cross section [pb]}\\\hline
    n & \multicolumn{2}{c|}{8} & \multicolumn{2}{c|}{9} 
    & \multicolumn{2}{c|}{10} & \multicolumn{2}{c|}{11} & \multicolumn{2}{c|}{12}\\
    $\sqrt{s}$ [GeV] & \multicolumn{2}{c|}{1500} & \multicolumn{2}{c|}{2000} 
    & \multicolumn{2}{c|}{2500} & \multicolumn{2}{c|}{3500} & \multicolumn{2}{c|}{5000}\\\hline
    Comix                          & 0&.755(3)  & 0&.305(2) & 0&.101(7) & 0&.057(5) & 0&.026(1) \\
    Ref.~\cite{Maltoni:2002mq}     & 0&.70(4)   & 0&.30(2)  & 0&.097(6) &  &        &  &        \\
    Ref.~\cite{Caravaglios:1998yr} & 0&.719(19) &  &        &  &        &  &        &  &        \\\hline
  \end{tabular}\end{center}\vspace*{5mm}\end{minipage}}{
  Cross sections for multi-gluon scattering at the 
  centre-of-mass energy $\sqrt{s}$, using the phase space cuts
  specified in Eq.~\eqref{eq:fabio_xsc}, compared to literature results.
  In parentheses the statistical error is stated in units of 
  the last digit of the cross section.\label{tab:fabio_xsc}}
\mytable{p}{
  \begin{minipage}{0.85\textwidth}\begin{center}
  \begin{tabular}{|p{2.45cm}
    |a{1.2cm}{0.8cm}|a{1.0cm}{1.0cm}|a{0.8cm}{1.2cm}|a{0.8cm}{1.2cm}|}\hline
    gg $\to$ ng\vphantom{\Large P} & \multicolumn{8}{c|}{Cross section [pb]}\\\hline
    n\vphantom{\Large P} & \multicolumn{2}{c|}{7} & \multicolumn{2}{c|}{8} 
    & \multicolumn{2}{c|}{9} & \multicolumn{2}{c|}{10}\\\hline
    Comix & 2703&(14) & 407&.0(36) & 66&.5(13) & 15&.2(26)\\\hline
    \end{tabular}\end{center}\vspace*{5mm}\end{minipage}}{
  Multi-gluon cross sections at the LHC with 
  $\sqrt{d}\ge 20\; {\rm GeV}$ and $d$ defined as in Ref.~\cite{Blazey:2000qt}, 
  except that $\Delta R_{ij}^2\to 2\,\sbr{\,\cosh\Delta\eta_{ij}-\cos\Delta\phi_{ij}\,}$.
  In parentheses the statistical error is stated in units 
  of the last digit of the cross section.\label{tab:gluon_xsc}}

\myfigure{p}{
  \begin{picture}(440,480)
    \put(6,0){\includegraphics[width=7.37cm]{figures/PS_6g20_t.eps}}
    \put(2,240){\includegraphics[width=7.52cm]{figures/PS_4g20_t.eps}}
    \put(230,0){\includegraphics[width=7.25cm]{figures/PS_7g20_t.eps}}
    \put(223,240){\includegraphics[width=7.7cm]{figures/PS_5g20_t.eps}}
    %\graphpaper[5](0,0)(500,500)
  \end{picture}\vspace*{5mm}}{
  Overall integration performance for multi-gluon scattering. 
  Upper panels display the Monte Carlo estimate of the cross section
  with the corresponding $1\sigma$ statistical error band as a function 
  of the total integration time. Lower panels show the relative 
  statistical error. 
  \Haag denotes the phase space integrator described in Ref.~\cite{Gleisberg:2008ft},
  applied on colour- and helicity-summed ME, generated using the CSW vertex rules.
  CSI denotes the integrator discussed in section~\ref{sec:csi}, applied on
  colour-sampled and helicity-summed ME's, generated using the CDBG recursion.
  Results for RAMBO were generated using colour- and helicity-sampled ME's
  form the CDBG recursion. Calculations have been performed on a 
  2.66~GHz Xeon$^{\texttrademark}$ CPU\label{fig:csi_conv_1}}
\myfigure{p}{
  \begin{picture}(440,230)
    \put(0,-1){\includegraphics[width=7.6cm]{figures/PS_8g20_t.eps}}
    \put(235,-2){\includegraphics[width=7.43cm]{figures/PS_9g20_t.eps}}
    %\graphpaper[5](0,0)(500,500)
  \end{picture}\vspace*{5mm}}{
  Overall integration performance for multi-gluon scattering, 
  continued from Fig.~\ref{fig:csi_conv_1}. Additionally, for the CSI a sampling 
  over helicity is considered, denoted by CSI(HS).\label{fig:csi_conv_2}}

\mytable{p}{
  \begin{tabular}{|c|c|}\hline
    Parameter & Value\\\hline
    \multicolumn{2}{c}{EW parameters in the $G_\mu$ scheme}\\\hline
    $G_F$ & $1.16639\times 10^{-5}$\\
    $\alpha_{QED}$ & 1/132.51\\
    $\sin^2\theta_W$ & 0.2222\\
    $M_W$ & 80.419 GeV\\
    $M_Z$ & 91.188 GeV\\
    $m_H$ & 120 GeV\\\hline
    \multicolumn{2}{c}{CKM matrix}\\\hline
    $V_{ud}, V_{cs}$ & 0.975\\\hline
    \multicolumn{2}{c}{QCD parameters}\\\hline
    PDF set & CTEQ6L1\\
    $\alpha_s$ & 0.130\\
    $\mu_F$, $\mu_R$ & $M_Z$\\
    jet, initial parton & $g$, $u$, $d$, $s$, $c$\\\hline
    \multicolumn{2}{c}{}\\
  \end{tabular}\hspace*{2ex}
  \begin{tabular}{|c|c|}\hline
    Parameter & Value\\\hline
    \multicolumn{2}{c}{Non-zero fermion masses (no evolution)}\\\hline
    $m_b$ & 4.7 GeV\\
    $m_t$ & 174.3 GeV\\
    $m_\tau$ & 1.777 GeV\\\hline
    \multicolumn{2}{c}{Widths (fixed width scheme)}\\\hline
    $\Gamma_W$ & 2.048 GeV\\
    $\Gamma_Z$ & 2.446 GeV\\
    $\Gamma_H$ & $3.7\times 10^{-3}$ GeV\\
    $\Gamma_t$ & 1.508 GeV\\
    $\Gamma_\tau$ & $2.36\times 10^{-12}$ GeV\\\hline
    \multicolumn{2}{c}{Cuts}\\\hline
    $p_{\perp,\,i}$ & $> 20$ GeV\\
    $|\eta_i|$ & $< 2.5$\\
    $\Delta R_{ij}$ & $> 0.4$\\\hline
    \multicolumn{2}{|c|}{no cuts on particles of $m>3$ GeV and $\nu_l$}\\\hline
  \end{tabular}\vspace*{2ex}}{
  Parameters for the MC4LHC comparison setup.\label{tab:mc4lhc_params}}
\newcommand{\nt}{$^\dagger$}
\mytable{p}{
\begin{minipage}{0.99\textwidth}
\begin{tabular}{|p{3cm}
  |a{0.7cm}{0.7cm}|a{0.5cm}{0.9cm}|a{0.4cm}{1.0cm}
  |a{0.3cm}{1.1cm}|a{0.3cm}{1.1cm}|a{0.2cm}{1.2cm}|a{0.2cm}{1.2cm}|}\hline
  $\sigma$ [$\mu$b]\vphantom{\Large P} & \multicolumn{14}{c|}{Number of jets}\\\hline
  $jets$\vphantom{\Large P} & \multicolumn{2}{c|}{2} & \multicolumn{2}{c|}{3} 
  & \multicolumn{2}{c|}{4} & \multicolumn{2}{c|}{5} & \multicolumn{2}{c|}{6} 
  & \multicolumn{2}{c|}{7} & \multicolumn{2}{c|}{8}\\\hline
  Comix    & 331&.0(4) & 22&.72(6) & 4&.95(2) & 1&.232(4) & 0&.352(1) & 0&.1133(5) & 0&.0369(3)\\
  ALPGEN   & 331&.7(3) & 22&.49(7) & 4&.81(1) & 1&.176(9) & 0&.330(1) &  &         &  &\\
  AMEGIC   & 331&.0(4) & 22&.78(6) & 4&.98(1) & 1&.238(4) &  &        &  &         &  &\\\hline
\end{tabular}\vspace*{3mm}
\begin{tabular}{|p{3cm}
  |a{0.7cm}{0.7cm}|a{0.5cm}{0.9cm}|a{0.3cm}{1.1cm}
  |a{0.3cm}{1.1cm}|a{0.2cm}{1.2cm}|a{0.2cm}{1.2cm}|a{0.2cm}{1.2cm}|}\hline
  $\sigma$ [$\mu$b]\vphantom{\Large P} & \multicolumn{14}{c|}{Number of jets}\\\hline
  $b\bar{b}$ + jets\vphantom{\Large P} & \multicolumn{2}{c|}{0} & \multicolumn{2}{c|}{1} 
  & \multicolumn{2}{c|}{2} & \multicolumn{2}{c|}{3} & \multicolumn{2}{c|}{4} 
  & \multicolumn{2}{c|}{5} & \multicolumn{2}{c|}{6}\\\hline
  Comix    & 471&.2(5) & 8&.83(2) & 1&.813(8) & 0&.459(2) & 0&.150(1) & 0&.0531(5) & 0&.0205(4)\\
  ALPGEN   & 470&.6(6) & 8&.83(1) & 1&.822(9) & 0&.459(2) & 0&.150(2) & 0&.053(1)  & 0&.0215(8)\\
  AMEGIC   & 470&.3(4) & 8&.84(2) & 1&.817(6) &  &        &  &        &  &         &  &\\\hline
\end{tabular}\vspace*{3mm}
\begin{tabular}{|p{3cm}
  |a{0.7cm}{0.7cm}|a{0.7cm}{0.7cm}|a{0.7cm}{0.7cm}
  |a{0.7cm}{0.7cm}|a{0.7cm}{0.7cm}|a{0.7cm}{0.7cm}|a{0.7cm}{0.7cm}|}\hline
  $\sigma$ [pb]\vphantom{\Large P} & \multicolumn{14}{c|}{Number of jets}\\\hline
  $t\bar{t}$ + jets\vphantom{\Large P} & \multicolumn{2}{c|}{0} & \multicolumn{2}{c|}{1} 
  & \multicolumn{2}{c|}{2} & \multicolumn{2}{c|}{3} & \multicolumn{2}{c|}{4} 
  & \multicolumn{2}{c|}{5} & \multicolumn{2}{c|}{6}\\\hline
  Comix    & 754&.8(8) & 745&(1) & 518&(1) & 309&.8(8) & 170&.4(7) &  89&.2(4) & 44&.4(4)\\
  ALPGEN   & 755&.4(8) & 748&(2) & 518&(2) & 310&.9(8) & 170&.9(5) &  87&.6(3) & 45&.1(8)\\
  AMEGIC & 754&.4(3) & 747&(1) & 520&(1) &    &      &    &      &    &      &   &\\\hline
\end{tabular}\vspace*{3mm}\end{minipage}}{
Cross sections $\sigma$ in the MC4LHC comparison~\cite{MC4LHC:2003aa} setup. In parentheses 
the statistical error is stated in units of the last digit of the cross section.
Note that for \Amegic and \Comix all subprocesses are considered, 
while \Alpgen is restricted to up to four quarks.\label{tab:mc4lhc_qcd}}
\mytable{p}{
\begin{minipage}{0.99\textwidth}
\begin{tabular}{|p{3cm}
  |a{0.9cm}{0.5cm}|a{0.9cm}{0.5cm}|a{0.7cm}{0.7cm}
  |a{0.7cm}{0.7cm}|a{0.5cm}{0.9cm}|a{0.5cm}{0.9cm}|a{0.5cm}{0.9cm}|}\hline
  $\sigma$ [pb]\vphantom{\Large P} & \multicolumn{14}{c|}{Number of jets}\\\hline
  $e^+\nu_e$ + jets\vphantom{\Large P} & \multicolumn{2}{c|}{0} & \multicolumn{2}{c|}{1} 
  & \multicolumn{2}{c|}{2} & \multicolumn{2}{c|}{3} & \multicolumn{2}{c|}{4} 
  & \multicolumn{2}{c|}{5} & \multicolumn{2}{c|}{6}\\\hline
  Comix    & 5434&(5) & 1274&(2)  & 465&(1) & 183&.0(6) & 77&.5(3) & 33&.8(1) & 14&.7(1)\\
  ALPGEN   & 5423&(9) & 1291&(13) & 465&(2) & 182&.8(8) & 75&.7(8) & 32&.5(2) & 13&.9(2)\\
  AMEGIC   & 5432&(5) & 1279&(2)  & 466&(2) & 185&.2(5) & 77&.3(4) &   &      &   &\\\hline
\end{tabular}\vspace*{3mm}
\begin{tabular}{|p{3cm}
  |a{0.9cm}{0.5cm}|a{0.9cm}{0.5cm}|a{0.7cm}{0.7cm}
  |a{0.7cm}{0.7cm}|a{0.7cm}{0.7cm}|a{0.5cm}{0.9cm}|a{0.5cm}{0.9cm}|}\hline
  $\sigma$ [pb]\vphantom{\Large P} & \multicolumn{14}{c|}{Number of jets}\\\hline
  $e^-\bar{\nu}_e$ + jets\vphantom{\Large P} & \multicolumn{2}{c|}{0} & \multicolumn{2}{c|}{1} 
  & \multicolumn{2}{c|}{2} & \multicolumn{2}{c|}{3} & \multicolumn{2}{c|}{4} 
  & \multicolumn{2}{c|}{5} & \multicolumn{2}{c|}{6}\\\hline
  Comix    & 3911&(4) & 1011&(2) & 362&(1) & 137&.1(3) & 54&.9(2) & 22&.4(1) & 9&.26(4)\\
  ALPGEN   & 3904&(6) & 1013&(2) & 364&(2) & 136&(1)   & 53&.6(6) & 21&.6(2) & 8&.7(1)\\
  AMEGIC   & 3903&(4) & 1012&(2) & 363&(1) & 137&.6(3) & 54&.8(6) &   &      &  &\\\hline
\end{tabular}\vspace*{3mm}
\begin{tabular}{|p{3cm}
  |a{0.7cm}{0.7cm}|a{0.7cm}{0.7cm}|a{0.5cm}{0.9cm}
  |a{0.5cm}{0.9cm}|a{0.5cm}{0.9cm}|a{0.5cm}{0.9cm}|a{0.5cm}{0.9cm}|}\hline
  $\sigma$ [pb]\vphantom{\Large P} & \multicolumn{14}{c|}{Number of jets}\\\hline
  $e^-e^+$ + jets\vphantom{\Large P} & \multicolumn{2}{c|}{0} & \multicolumn{2}{c|}{1} 
  & \multicolumn{2}{c|}{2} & \multicolumn{2}{c|}{3} & \multicolumn{2}{c|}{4} 
  & \multicolumn{2}{c|}{5} & \multicolumn{2}{c|}{6}\\\hline
  Comix    & 723&.5(4) & 187&.9(3) & 69&.7(2) & 27&.14(7) & 11&.09(4) & 4&.68(2) & 2&.02(2)\\
  ALPGEN   & 723&.4(9) & 188&.3(3) & 69&.9(3) & 27&.2(1)  & 10&.95(5) & 4&.6(1)  & 1&.85(1)\\
  AMEGIC   & 723&.0(8) & 188&.2(3) & 69&.6(2) & 27&.21(6) & 11&.1(1)  &  &       &  &\\\hline
\end{tabular}\vspace*{3mm}
\begin{tabular}{|p{3cm}
  |a{0.9cm}{0.5cm}|a{0.7cm}{0.7cm}|a{0.7cm}{0.7cm}
  |a{0.7cm}{0.7cm}|a{0.7cm}{0.7cm}|a{0.5cm}{0.9cm}|a{0.5cm}{0.9cm}|}\hline
  $\sigma$ [pb]\vphantom{\Large P} & \multicolumn{14}{c|}{Number of jets}\\\hline
  $\nu_e\bar{\nu}_e$ + jets\vphantom{\Large P} & \multicolumn{2}{c|}{0} & \multicolumn{2}{c|}{1} 
  & \multicolumn{2}{c|}{2} & \multicolumn{2}{c|}{3} & \multicolumn{2}{c|}{4} 
  & \multicolumn{2}{c|}{5} & \multicolumn{2}{c|}{6}\\\hline
  Comix    & 3266&(3) & 715&.9(8) & 266&.6(7) & 105&.0(3) & 44&.4(2) & 19&.11(7) & 8&.30(7)\\
  ALPGEN   & 3271&(1) & 717&.4(5) & 267&.4(4) & 105&.4(2) & 43&.7(2) & 18&.68(8) & 7&.88(5)\\
  AMEGIC   & 3270&(1) & 717&.3(7) & 266&.3(6) & 105&.4(3) & 44&.3(5) &   &       &  &\\\hline
\end{tabular}\vspace*{3mm}
\begin{tabular}{|p{3cm}
  |a{0.5cm}{0.9cm}|a{0.5cm}{0.9cm}|a{0.5cm}{0.9cm}
  |a{0.5cm}{0.9cm}|a{0.5cm}{0.9cm}|a{0.5cm}{0.9cm}|a{0.5cm}{0.9cm}|}\hline
  $\sigma$ [pb]\vphantom{\Large P} & \multicolumn{14}{c|}{Number of jets}\\\hline
  $\gamma\gamma$ + jets\vphantom{\Large P} & \multicolumn{2}{c|}{0} & \multicolumn{2}{c|}{1} 
  & \multicolumn{2}{c|}{2} & \multicolumn{2}{c|}{3} & \multicolumn{2}{c|}{4} 
  & \multicolumn{2}{c|}{5} & \multicolumn{2}{c|}{6}\\\hline
  Comix    & 45&.64(5) & 25&.23(6) & 18&.57(6) & 9&.64(4) & 4&.65(2) & 2&.07(2) & 0&.88(3)\\
  AMEGIC & 45&.66(3) & 25&.41(6) & 18&.81(7) & 9&.82(3) &  &       &  &       &  &\\\hline
\end{tabular}\vspace*{3mm}\end{minipage}}{
Cross sections $\sigma$ in the MC4LHC comparison~\cite{MC4LHC:2003aa} setup. In parentheses 
the statistical error is stated in units of the last digit of the cross section.
Note that for \Amegic and \Comix all subprocesses are considered, 
while \Alpgen is restricted to up to four quarks.\label{tab:mc4lhc_ewqcd}}

\mytable{p}{
\begin{minipage}{15.5cm}
\begin{tabular}{|p{4cm}
  |a{0.7cm}{0.7cm}|a{0.5cm}{0.9cm}|a{0.5cm}{0.9cm}
  |a{0.3cm}{1.1cm}|a{0.3cm}{1.1cm}|a{0.3cm}{1.1cm}|}\hline
  $\sigma$ [nb]\vphantom{\Large P} & \multicolumn{12}{c|}{Number of jets}\\\hline
  $\gamma$ + jets\vphantom{\Large P} & \multicolumn{2}{c|}{1} 
  & \multicolumn{2}{c|}{2} & \multicolumn{2}{c|}{3} & \multicolumn{2}{c|}{4} 
  & \multicolumn{2}{c|}{5} & \multicolumn{2}{c|}{6}\\\hline
  Comix    & 89&.5(2) & 19&.65(6) & 7&.52(3) & 2&.664(8) & 1&.000(5) & 0&.387(2)\\
  AMEGIC   & 89&.6(1) & 19&.60(5) & 7&.59(2) & 2&.64(2)  &  &        &  &\\\hline
\end{tabular}\vspace*{3mm}
\begin{tabular}{|p{4cm}
  |a{0.5cm}{0.9cm}|a{0.5cm}{0.9cm}|a{0.5cm}{0.9cm}
  |a{0.5cm}{0.9cm}|a{0.5cm}{0.9cm}|a{0.3cm}{1.1cm}|}\hline
  $\sigma$ [pb]\vphantom{\Large P} & \multicolumn{12}{c|}{Number of jets}\\\hline
  $e^-\bar{\nu}_e$ + $b\bar{b}$ + jets\vphantom{\Large P} & 
  \multicolumn{2}{c|}{0} & \multicolumn{2}{c|}{1} 
  & \multicolumn{2}{c|}{2} & \multicolumn{2}{c|}{3} 
  & \multicolumn{2}{c|}{4} & \multicolumn{2}{c|}{5}\\\hline
  Comix    & 9&.40(2) & 9&.81(3) & 6&.82(5) & 4&.32(4) & 2&.47(2) & 1&.28(2)\\
  ALPGEN   & 9&.34(4) & 9&.85(6) & 6&.82(6) & 4&.18(7) & 2&.39(5) &  &\\
  AMEGIC   & 9&.37(1) & 9&.86(2) & 6&.98(3) & 4&.31(6) &  &       &  &\\\hline
\end{tabular}\vspace*{3mm}
\begin{tabular}{|p{4cm}
  |a{0.5cm}{0.9cm}|a{0.5cm}{0.9cm}|a{0.5cm}{0.9cm}
  |a{0.3cm}{1.1cm}|a{0.3cm}{1.1cm}|a{0.3cm}{1.1cm}|}\hline
  $\sigma$ [pb]\vphantom{\Large P} & \multicolumn{12}{c|}{Number of jets}\\\hline
  $e^-e^+$ + $b\bar{b}$ + jets\vphantom{\Large P} & 
  \multicolumn{2}{c|}{0} & \multicolumn{2}{c|}{1} 
  & \multicolumn{2}{c|}{2} & \multicolumn{2}{c|}{3} 
  & \multicolumn{2}{c|}{4} & \multicolumn{2}{c|}{5}\\\hline
  Comix    & 18&.90(3) & 6&.81(2) & 3&.07(3) & 1&.536(9) & 0&.763(6) & 0&.37(1)\\
  ALPGEN   & 18&.95(8) & 6&.80(3) & 2&.97(2) & 1&.501(9) & 0&.78(1)  &  &\\
  AMEGIC   & 18&.90(2) & 6&.82(2) & 3&.06(4) &  &        &  &        &  &\\\hline
\end{tabular}\vspace*{3mm}\end{minipage}}{
Cross sections $\sigma$ in the MC4LHC comparison~\cite{MC4LHC:2003aa} setup. In parentheses 
the statistical error is stated in units of the last digit of the cross section.
Note that for \Amegic and \Comix all subprocesses are considered, 
while \Alpgen is restricted to up to four quarks.\label{tab:mc4lhc_vb}}

\mytable{p}{
\begin{minipage}{0.99\textwidth}\begin{center}
\begin{tabular}{|p{2.5cm}
  |a{0.4cm}{1.4cm}|a{0.4cm}{1.6cm}|}\hline
  $\sigma$ [nb] \vphantom{\Large P} & \multicolumn{4}{c|}{Number of jets $n$}\\\hline
  QCD jets\vphantom{{\Large P}} & \multicolumn{2}{c|}{7} & 
  \multicolumn{2}{c|}{8}\\\hline
  $gg\to ng$\vphantom{{\Large P}} & 49&.1(4)     & 14&.2(3)      \\
  $gg\to (n-2)g\,2q$ &              17&.0(1)     &  6&.0(1)      \\
  $gg\to (n-4)g\,4q$ &               1&.69(1)    &  0&.74(5)     \\
  $gg\to (n-6)g\,6q$ &               0&.0401(5)  &  0&.0297(8)   \\
  $gg\to 8q$         &               -&          &  0&.000158(5) \\
  $gq\to (n-1)g\,1q$ &              30&.5(2)     &  9&.9(2)      \\
  $gq\to (n-3)g\,3q$ &               8&.46(6)    &  3&.38(6)     \\
  $gq\to (n-5)g\,5q$ &               0&.565(7)   &  0&.332(8)    \\
  $gq\to (n-7)g\,7q$ &               0&.00501(6) &  0&.0067(2)   \\
  $qq\to ng$         &               0&.0209(1)  &  0&.0067(1)   \\
  $qq\to (n-2)g\,2q$ &               4&.97(4)    &  1&.84(3)     \\
  $qq\to (n-4)g\,4q$ &               1&.044(9)   &  0&.477(9)    \\
  $qq\to (n-6)g\,6q$ &               0&.0374(3)  &  0&.0291(5)   \\
  $qq\to 8q$         &               -&          &  0&.000223(4) \\\hline
\end{tabular}\hspace*{5mm}
\begin{tabular}{|p{3.3cm}
  |a{0.4cm}{1.2cm}|a{0.3cm}{1.5cm}|}\hline
  $\sigma$ [pb] \vphantom{\Large P} & \multicolumn{4}{c|}{Number of jets $n$}\\\hline
  $e^+\nu_e$ + QCD jets\vphantom{{\Large P}} & \multicolumn{2}{c|}{5} & 
  \multicolumn{2}{c|}{6}\\\hline
  $qq\to e^+\nu_e\,ng$\vphantom{{\Large P}} &  0&.256(2)  & 0&.0768(6)  \\
  $qq\to e^+\nu_e\,(n-2)g\,2q$ &               6&.49(3)   & 2&.92(3)    \\
  $qq\to e^+\nu_e\,(n-4)g\,4q$ &               0&.591(3)  & 0&.449(8)   \\
  $qq\to e^+\nu_e\,6q$         &               -&         & 0&.00640(7) \\
  $gq\to e^+\nu_e\,(n-1)g\,1q$ &              20&.0(1)    & 8&.21(8)    \\
  $gq\to e^+\nu_e\,(n-3)g\,3q$ &               4&.03(2)   & 2&.14(2)    \\
  $gq\to e^+\nu_e\,(n-5)g\,5q$ &               0&.0741(4) & 0&.094(1)   \\
  $gg\to e^+\nu_e\,(n-2)g\,2q$ &               2&.13(1)   & 0&.775(5)   \\
  $gg\to e^+\nu_e\,(n-4)g\,4q$ &               0&.1817(9) & 0&.1058(7)  \\
  $gg\to e^+\nu_e\,6q$         &               -&         & 0&.001403(7)\\\hline
\multicolumn{5}{c}{}\\
\multicolumn{5}{c}{}\\
\multicolumn{5}{c}{}\\
\multicolumn{5}{c}{}\\
\end{tabular}\vspace*{5mm}
\end{center}\end{minipage}}{
Subprocess cross sections $\sigma$ in the MC4LHC comparison~\cite{MC4LHC:2003aa} 
setup. In parentheses the statistical error is stated in units of the last digit 
of the cross section.\label{tab:mc4lhc_sub}}

\mytable{p}{
\begin{minipage}{0.99\textwidth}
\begin{tabular}{|p{3cm}
  |a{0.5cm}{0.9cm}|a{0.5cm}{0.9cm}|a{0.5cm}{0.9cm}
  |a{0.5cm}{0.9cm}|a{0.5cm}{0.9cm}|a{0.5cm}{0.9cm}|a{0.5cm}{0.9cm}|}\hline
  efficiency\vphantom{\Large P} & \multicolumn{14}{c|}{Number of jets}\\\hline
  jets\vphantom{{\Large P}} & \multicolumn{2}{c|}{2} & 
  \multicolumn{2}{c|}{3} & \multicolumn{2}{c|}{4} & \multicolumn{2}{c|}{5} & 
  \multicolumn{2}{c|}{6} & \multicolumn{2}{c|}{7} & \multicolumn{2}{c|}{8}\\\hline
  $\varepsilon=10^{-3}$ \vphantom{{\Large P}} & 
  9.3&$\cdot10^{-2}$ & 7.8&$\cdot10^{-3}$ & 2.1&$\cdot10^{-3}$ & 
  7.0&$\cdot10^{-4}$ & 3.6&$\cdot10^{-4}$ & 1.3&$\cdot10^{-4}$ & 6.1&$\cdot10^{-5}$ \\
  $\varepsilon=10^{-6}$ \vphantom{{\Large P}} & 
  3.1&$\cdot10^{-2}$ & 3.8&$\cdot10^{-3}$ & 1.5&$\cdot10^{-3}$ & 
  4.3&$\cdot10^{-4}$ & 2.4&$\cdot10^{-4}$ & 9.9&$\cdot10^{-5}$ & 5.8&$\cdot10^{-5}$ \\
%  AMEGIC++ & 2.5&$\cdot10^{-2}$ & 4.5&$\cdot10^{-3}$ & 5.0&$\cdot10^{-3}$ & 
%  9.4&$\cdot10^{-4}$ & & & & & & \\
  \hline
\end{tabular}\vspace*{5mm}
\begin{tabular}{|p{3cm}
  |a{0.5cm}{0.9cm}|a{0.5cm}{0.9cm}|a{0.5cm}{0.9cm}
  |a{0.5cm}{0.9cm}|a{0.5cm}{0.9cm}|a{0.5cm}{0.9cm}|a{0.5cm}{0.9cm}|}\hline
  efficiency\vphantom{\Large P} & \multicolumn{14}{c|}{Number of jets}\\\hline
  $e^+\nu_e$ + jets\vphantom{{\Large P}} & \multicolumn{2}{c|}{0} & 
  \multicolumn{2}{c|}{1} & \multicolumn{2}{c|}{2} & \multicolumn{2}{c|}{3} & 
  \multicolumn{2}{c|}{4} & \multicolumn{2}{c|}{5} & \multicolumn{2}{c|}{6}\\\hline
  $\varepsilon=10^{-3}$ \vphantom{{\Large P}} & 
  1.5&$\cdot10^{-1}$ & 2.4&$\cdot10^{-2}$ & 9.1&$\cdot10^{-3}$ & 
  2.0&$\cdot10^{-3}$ & 6.7&$\cdot10^{-4}$ & 1.9&$\cdot10^{-4}$ & 3.1&$\cdot10^{-5}$ \\
  $\varepsilon=10^{-6}$ \vphantom{{\Large P}} & 
  1.6&$\cdot10^{-2}$ & 4.5&$\cdot10^{-3}$ & 3.3&$\cdot10^{-3}$ & 
  1.2&$\cdot10^{-3}$ & 4.3&$\cdot10^{-4}$ & 1.3&$\cdot10^{-4}$ & 2.8&$\cdot10^{-5}$ \\
%  AMEGIC++ & 1.1&$\cdot10^{-2}$ & 6.4&$\cdot10^{-3}$ & 2.6&$\cdot10^{-3}$ & 
%  4.9&$\cdot10^{-4}$ & & & & & & \\
  \hline
\end{tabular}\vspace*{5mm}
\end{minipage}}{
Efficiencies for processes in the MC4LHC comparison~\cite{MC4LHC:2003aa} 
setup.\label{tab:mc4lhc_effs}}

%% file: text/conclusions.tex
We have presented the new matrix element generator \Comix,
based on the colour dressed Berends-Giele
recursive relations and two new methods for phase space generation. 
We have analysed the performance of the new generator and compared
the respective results to other ME generators. We find that the
new algorithms perform very well and we obtain promising results
for large multiplicity processes. \Comix can therefore be considered
an excellent supplementary generator for large multiplicities,
which is especially helpful in the context of a matrix element -
parton shower merging. The treatment of colour in \Comix makes the
algorithm well suited for such an interface, since the colour 
structure of the matrix element does not need to be guessed from
the kinematics, it is rather fixed on a point by point basis.
A corresponding publication is forthcoming~\cite{ckkw}.

%% file: text/acknowledgements.tex
We like to thank Claude Duhr, Frank Krauss and Fabio Maltoni
for fruitful discussions and their comments on the manuscript.
Special thanks for technical support go to Jonathan Ferland, 
Phil Roffe, Graeme Stewart and the ScotGrid~\cite{ScotGrid:2008aa} 
Tier 2 sites Durham and Glasgow. We thank Steffen Schumann
for providing comparison results from \Amegic and Michelangelo
Mangano for results from \Alpgen.
TG's research was supported by the US Department of Energy,
contract DE-AC02-76SF00515.
SH thanks the HEPTOOLS Marie Curie Research Training Network 
(contract number MRTN-CT-2006-035505) for an Early Stage 
Researcher position. Support from MCnet (contract number 
MRTN-CT-2006-035606) is acknowledged.